\documentclass{JHEP3}
\usepackage{graphics}
\usepackage{amsmath,amsfonts}
\usepackage{latexsym}

\title{Revamped Braneworld Gravity}

\author{Ruoyu Bao$^\dag$, Marcela Carena$^\ddag$, 
Joseph Lykken$^{\dag\;\ddag}$, Minjoon Park$^\dag$ 
and Jos\'e Santiago$^\ddag$ 
\thanks{\texttt{rbao@theory.uchicago.edu, carena@fnal.gov, 
lykken@fnal.gov, mpark@uchicago.edu, jsantiag@fnal.gov }}\\
$^\dag$Enrico Fermi Institute and 
Department of Physics, The University of Chicago,\\
5640 South Ellis Ave., Chicago, IL 60637, USA\\
$^\ddag$Fermi National Accelerator Laboratory,
P.O. Box 500, Batavia, IL 60510, USA}

\date{}

\abstract{Gravity in five-dimensional braneworld backgrounds often
exhibits problematic features, including kinetic ghosts, strong
coupling, and the vDVZ discontinuity. These problems are an obstacle
to producing and analyzing braneworld models with interesting and
potentially observable modifications of 4d gravity. We examine these problems
in a general $AdS_5/AdS_4$ setup with two branes and localized
curvature from arbitrary
brane kinetic terms. We use the interval approach and an explicit
``straight'' gauge-fixing. We compute the complete quadratic
gauge-fixed effective 4d action, as well as the leading cubic
order corrections. We compute the exact Green's function for gravity
as seen on the brane. In the full parameter space, we exhibit
the regions which avoid kinetic ghosts and tachyons. We give a general formula
for the strong coupling scale, \textit{i.e.} the energy scale at which
the linearized treatment of gravity breaks down, for relevant
regions of the parameter space. We show how the vDVZ discontinuity 
can be naturally but nontrivially avoided by ultralight graviton modes.
We present a direct comparison of warping versus
localized curvature in terms of their effects on graviton mode
couplings. We exhibit the first example of
DGP-like crossover behavior in a general warped setup.

}

\preprint{\normalsize\rm FERMILAB-Pub-05/510-T\\EFI-05-20}

\begin{document}

\newcommand{\be}{\begin{eqnarray}}
\newcommand{\ee}{\end{eqnarray}}
\newcommand{\bea}{\begin{eqnarray}}
\newcommand{\eea}{\end{eqnarray}}
\newcommand{\nn}{\nonumber}
\newcommand{\bd}{\begin{displaymath}}
\newcommand{\ed}{\end{displaymath}}

\baselineskip=16pt
\newcommand{\gsim}{\ \rlap{\raise 2pt\hbox{$>$}}{\lower 2pt \hbox{$\sim$}}\ }
\newcommand{\lsim}{\ \rlap{\raise 2pt\hbox{$<$}}{\lower 2pt \hbox{$\sim$}}\ }

\def\etal{{\it et al.}}
\def\ie{{\it i.e.}}
\def\eg{{\it e.g.}}
\def\tn{\tilde\nabla}
\def\hn{\hat\nabla}
\def\pa{\partial}
\def\gto#1{\;\to \kern -18pt \lower 5pt\hbox{$\scriptstyle #1$}\;}



\def\VEV#1{\left\langle{ #1} \right\rangle}
\def\bra#1{\left\langle{ #1} \right|}
\def\ket#1{\left| {#1} \right\rangle}
\def\vev#1{\langle #1 \rangle}
\def\norm#1{\left\langle{ #1} \vert {#1} \right\rangle}


\def\One{{\bf 1}}
\def\hc{{\mbox{\rm h.c.}}}
\def\tr{{\mbox{\rm tr}}}
\def\half{\frac{1}{2}}
\def\thalf{\frac{3}{2}}

\def\Dslash{\not{\hbox{\kern-4pt $D$}}}
\def\dslash{\not{\hbox{\kern-2pt $\del$}}}


\newpage

\section{Introduction}

Gravity in five-dimensional braneworld backgrounds exhibits
features which, from a 4d point of view, are both novel and
surprising. Examples include:
\begin{itemize}
\item Kaluza-Klein (KK) graviton modes whose couplings
to brane matter are only TeV suppressed \cite{RSI}, 
and thus potentially detectable as resonances at 
the LHC \cite{Davoudiasl:1999jd}--\cite{Bao:2005ni}.
\item Extra 4d scalar modes (radions) whose couplings to brane
matter are TeV suppressed but otherwise like those of 
gravity \cite{Goldberger:1999wh}--\cite{Das:2002me}.
\item Continuum KK graviton modes which are potentially
detectable at the LHC \cite{LR},\cite{Lykken:2000wz}.
\item KK graviton modes whose couplings to brane matter
are suppressed by warping \cite{RSII},\cite{KR}.
\item KK graviton modes whose couplings to brane matter
are suppressed by localized curvature (\textit{i.e.}
brane kinetic terms for gravity) \cite{DGP}.
\item 4d effective theories with no massless 
graviton \cite{KR},\cite{DGP}.
\item 4d effective theories with a massless graviton
and an additional ultralight KK graviton, whose mass
goes to zero in a well-defined 
parametric limit \cite{KMP}--\cite{Pthesis}.
\item A crossover scale, such that KK gravitons with
mass below this scale have unsuppessed couplings to
brane matter, but heavier KK modes have suppressed couplings.
This leads to modifications of 4d gravity which appear
in the infrared rather than the 
ultraviolet \cite{DGP},\cite{Dvali:2001gm}--\cite{Kiritsis:2002ca}.
Such ``DGP'' scenarios could have relevance 
to cosmology \cite{Deffayet:2000uy}--\cite{Lue:2005ya}.
\end{itemize}

What is truely remarkable is that these novel features of gravity
do not require the assumption of any new physics beyond a single
extra dimension and the existence of co-dimension one branes. The
above results are derived in low energy linearized effective descriptions
which require no assumptions about quantum gravity, string theory,
or the existence of exotic matter.

Before we celebrate too loudly, however, we should confront a number
of troubling issues common to most or all braneworld gravity models:
\begin{itemize}
\item Are these low energy effective descriptions really
just special cases of 5d General Relativity? This is not obvious
since most models invoke orbifold backgrounds, and since they
also ignore the problem of ultraviolet matching to 
\textit{e.g.} the microscopic features of the branes.
\item What are the physical degrees of freedom in these various
models? This is a difficult question to answer rigorously,
since the usual gauge-fixing choices (\textit{e.g.} harmonic gauge)
are not suitable for setups with more than one brane.
\item Are these models stable? Models with massless radions are
at best marginally stable. 
Some simple setups \cite{GRS}--\cite{Chacko:2003yp}
have radions which
are kinetic ghosts, \textit{i.e.} their kinetic terms have the
wrong sign. Kinetic ghosts indicate an instability in the model.
As we will see in this paper, graviton modes can also be kinetic
ghosts, and some kinetic ghosts are also tachyons. Although kinetic ghosts
may be useful for some purposes \cite{Arkani-Hamed:2003uy}--\cite{Pospelov},
their presence cannot safely be ignored.
\item Under what conditions do ultralight graviton modes
mimic 4d gravity, avoiding the famous vDVZ discontinuity \cite{vDVZ}?
\item At what scale does the low energy linearized approximation
break down? Perturbation theory involves more input parameters
than just the 5d gravitational coupling $M$; it would not be surprising
if in certain limits of these additional parameters the low energy
effective theory breaks down at a scale much lower than $M$.
Indeed this is precisely what happens in the simplest models
containing ultralight graviton modes or crossover behavior,
where strong coupling sets in at a scale parametrically much
smaller than $M$.
\end{itemize}  

In a recent investigation \cite{clp} three of us resolved the first two issues
listed above. We recast the standard braneworld models in
the ``interval picture'', where orbifolding is replaced by
intervals with boundaries. We showed that braneworld gravity
has a well-defined action principle only if we extend General Relativity
to include ``brane-boundary equations'' which supplement the usual
bulk Einstein equations. We also showed how to rigorously extract the
physical degrees of freedom, by introducing a class of ``straight''
gauges suitable for braneworld analysis.

The purpose of this paper is to address the remaining three issues
listed above, and thus to revamp the promising field of braneworld
gravity. In section 2 we derive the general 4d quadratic effective
action for a general setup with two branes, including warping and
localized curvature. This action is gauge-fixed to just the physical
degrees of freedom, using a straight gauge. In section 3 we use
this effective action to demonstrate the presence or absence of
kinetic ghosts and/or tachyons according to various choices of 
the input parameters.
This analysis is a generalization and improvement of 
earlier attempts \cite{Padilla}, \cite{Dubovsky:2003pn}.
We find that both the radion and the
graviton zero mode can be kinetic ghosts in certain regions 
of the parameter space.

In section 4 we address the question of strong coupling. By computing
the relevant part of the 4d cubic effective action, we give a general
formula to determine the strong coupling scale for the radion.
We show that the strong coupling scale becomes small in a
DGP-like limit.
In section 5 we give an exact expression for the straight gauge graviton
Green's function on the brane. This allows us to show how the
vDVZ discontinuity is avoided in warped models with ultralight
graviton modes. In a DGP-like limit, our straight gauge 
graviton Green's function
does not have any diverging tensor structures such as those that
appear in DGP; the potential breakdown of linearized gravity in this
limit is entirely due to the radion.

Finally, in section 6 we use the exact Green's function to study
the couplings of KK gravitons to brane matter. In the special
case of the Karch-Randall setup, we regain the recent results of
Kaloper and Sorbo \cite{Kaloper:2005wq}.
We focus on models with an infinite extra dimension and
no massless graviton.
Localized curvature and warping both have
the effect of making an ultralight graviton from the first
massive KK mode. We compare these two effects in our models,
and show that the warping effect is more efficient than
localized curvature in creating a simulacrum of 4d gravity.

In section 6 we exhibit explict crossover
behavior for models in our general setup. In these models
the couplings of the KK graviton modes to brane matter are
unsuppressed up to a mass scale $1/r_c$, and are
highly suppressed for modes heavier than this scale.
This behavior is the most interesting phenomenological
feature of DGP braneworld gravity; here we can study it
in a general warped framework.

\section{The 4d quadratic effective action}
In \cite{clp}, a brane world gravity theory with the action
\bea
\label{eqn:ouraction}
S &=&  
\int d^4x\,
\left(  \int^{L^-}_{0^+} dy 
+ \int^{0^-}_{-L^+} dy \right)
\sqrt{-G} \Big(2M^3 R - \Lambda\Big) \nn\\
&&+ \sum_i \int_{y=y_i} d^4 x 
\sqrt{-g^{(i)}} (2M_i^2 \tilde{\cal R}^{(i)} - V_i) 
+ 4M^3 \oint_{\partial \cal M} K \,.
\eea
was analyzed. This action represents a general warped
gravity setup with codimension one branes, written
in the interval picture. In (\ref{eqn:ouraction}) $M$
is the 5d Planck scale, $\Lambda = -24M^3k^2$
is the bulk cosmological constant giving a bulk curvature $k$,
the $M_i$ are the coefficients of brane-localized
curvatures $\tilde{\cal R}^{(i)}$,
the $V_i$ are brane tensions, and $K$ is the
extrinsic curvature of the Gibbons-Hawking boundary term.

Upon linearization, $G_{MN} = G^{\bf 0}_{MN} + h_{MN}$, 
the $AdS_5/AdS_4$ background solution is
\be\label{eqn:metric}
G^{\rm\bf 0}_{MN} = \begin{pmatrix}
g^{\rm\bf 0}_{\mu\nu} & 0 \\ 0 & 1 \end{pmatrix} \,,
\ee
where
\be
g^{\rm\bf 0}_{\mu\nu} = \frac{a(y)^2}{(1 
- \frac{H^2 x^2}{4})^2} \eta_{\mu\nu}\,,
\ee
with 
\be
a(y) = \frac{\cosh k(y-y_0)}{\cosh k y_0}\;,\quad & 0 < y < L\;, 
\ee
$\eta_{\mu\nu} = {\rm diag}(-1, 1, 1, 1)$ 
and $x^2 = \eta_{\mu\nu} x^\mu x^\nu$. Hereafter, we will omit the 
superscript ${}^{\bf 0}$. 

\begin{figure}
  \begin{center}
    \resizebox{11cm}{!}{\includegraphics{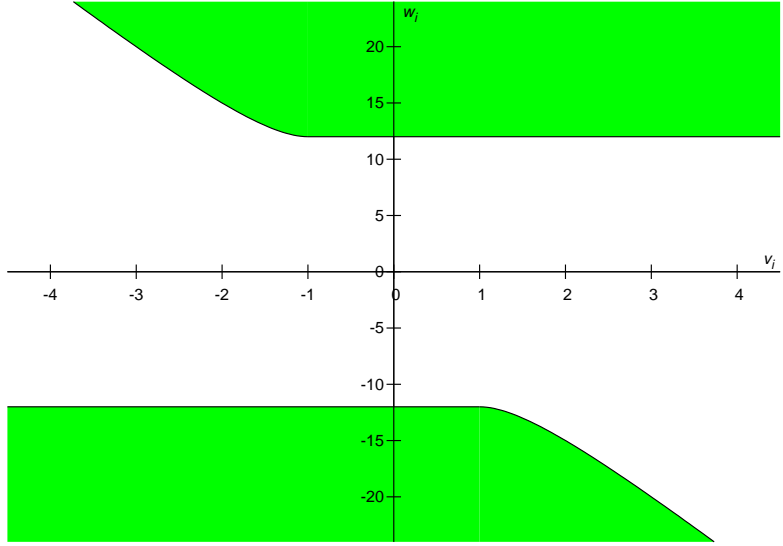}}
  \end{center}
  \caption{The unshaded area is where $-1<T_{i}^+<1$. On the boundary, 
the curved lines are described by 
  $w_i = -6 v_i - \frac{6}{v_i}$, and the straight lines are $w_i = \pm12$.}
  \label{talg}
\end{figure}

It is extremely convenient to trade $M_i$ and $V_i$ for
dimensionless input parameters $v_i$ and $w_i$ defined by
\be
v_i = k M_i^2/M^3 \; ,\qquad w_i = V_i / 2kM^3 \; .
\ee
The brane separation $L$ and the warping parameter $y_0$ are
then determined from these inputs by the relations 
\be\label{eqn:solfort}
T_i &=& \frac{w_i}{12} + \frac{v_i}{2}(1-T_i^2) \,,
\ee
or
\be
T_{i}^\pm = \frac{1}{v_i} 
\Big(-1 \pm \sqrt{1 + \frac{1}{6}w_i v_i+ v_i^2} \,\Big)\,,
\ee
with
\be
T_0 = \tanh k y_0 \; , \qquad T_L = \tanh k (L-y_0) \; .
\ee 
The $AdS_4$ inverse radius of curvature $H$ is given by
\be
H = \frac{k}{{\rm cosh}\,ky_0} \; .
\ee
Solving $-1<T_{i}^+<1$, Figure \ref{talg} shows the
range of input parameters which gives $AdS_5/AdS_4$ backgrounds.

Using this background solution, we eliminate gauge degrees
of freedom and find the following 4d physical degrees of freedom:
\begin{enumerate}
\item{At the massive level, we have a KK tower of massive spin-2 particles 
with 5 degrees of freedom each.
We can decompose the graviton explicitly into this tower of
massive modes:
\be\label{eqn:massiveh}
h_{\mu\nu} = \sum_{q} b^{(q)}_{\mu\nu}
= \sum_{q}{\mathcal Y}^{(q)}(y) B^{(q)}_{\mu\nu}(x)\,,
\ee
where $q$ labels the mass, $q =m^2/H^2$, and
\bea\label{eqn:massiveysol}
{\mathcal Y}^{(q)}(y) &=& P_{(-1+\sqrt{9+ 4q})/2}^{-2}(z) 
- \frac{a^{(q)}_0}{b^{(q)}_0} Q_{(-1+\sqrt{9+ 4q})/2}^2(z) \nn\\
&=& P_{(-1+\sqrt{9+ 4q})/2}^{-2}(z) 
- \frac{a^{(q)}_L}{b^{(q)}_L} Q_{(-1+\sqrt{9+ 4q})/2}^2(z) \,;
\eea
here $z=\tanh k(y-y_0)$ and the $P$'s and $Q$'s are associated 
Legendre functions.
The mass spectrum of modes is determined by solving the determinant equation, 
\bea\label{eqn:deteq}
a_0 b_L - a_L b_0  = 0\; ,
\eea
with
\bea\label{eqn:ourasandbs}
\hspace{-15pt} a_0 &=& 
\{v_0 q (1-T_0^2) + (3+\sqrt{9+ 4q})T_0 \} 
\,P_{(-1+\sqrt{9+ 4q})/2}^{-2}(-T_0)  \nn\\
&&\hspace{-15pt} \qquad +(-5+\sqrt{9+ 4q}) \,
P_{(-3+\sqrt{9+ 4q})/2}^{-2}(-T_0) \nn\\
\hspace{-15pt} &=& \sqrt{1-T_0^2}\,\{v_0 q \sqrt{1-T_0^2} 
\,P_{(-1+\sqrt{9+ 4q})/2}^{-2}(-T_0) 
- 2 P_{(-1+\sqrt{9+ 4q})/2}^{-1}(-T_0) \} \,, \nn\\
\hspace{-15pt} b_0 &=& q\sqrt{1-T_0^2}\, \{v_0 \sqrt{1-T_0^2}
\,Q_{(-1+\sqrt{9+ 4q})/2}^2(-T_0) 
+ 2 Q_{(-1+\sqrt{9+ 4q})/2}^1(-T_0) \} \,, \\
\hspace{-15pt} a_L &=& \sqrt{1-T_L^2} \{v_L q \sqrt{1-T_L^2}
\,P_{(-1+\sqrt{9+ 4q})/2}^{-2}(T_L) 
+ 2 P_{(-1+\sqrt{9+ 4q})/2}^{-1}(T_L) \} \,, \nn\\
\hspace{-15pt} b_L &=& q\sqrt{1-T_L^2}\,\{ v_L \sqrt{1-T_L^2}
\,Q_{(-1+\sqrt{9+ 4q})/2}^2(T_L) 
- 2 Q_{(-1+\sqrt{9+ 4q})/2}^1(T_L) \} \,. \nn
\eea
}
\item{At the massless level, there is a massless 
spin-2 particle (graviton) $\beta_{\mu\nu}$ with 2 degrees of freedom
and a 4d scalar (radion) $\psi(x)$. Thus $h_{\mu\nu}$ is 
\be\label{eqn:masslessh}
h_{\mu\nu} = \beta_{\mu\nu} 
+ a^2 {\mathcal Y}_1(y) \tn_\mu \tn_\nu \psi 
+ g_{\mu\nu} {\mathcal Y}_2(y) \psi \,. 
\ee
The $y$-dependence of $\beta_{\mu\nu}$ is determined to be
\be
\beta_{\mu\nu} = a^2(y) B_{\mu\nu}(x)\,. 
\ee}
The remaining $y$-dependence is in the ${\mathcal Y}_i$'s, which
are given by
\bea
{\mathcal Y}_1(y) = c (1-z)^2 + d z - {\mathfrak F}\,,\quad
{\mathcal Y}_2(y) = -H^2 c (1-z)^2 -H^2 d z + \frac{a'}{a} {\mathcal F}\,,
\eea
where
\bea\label{eqn:csol}
c &=& \frac{k}{H^2} \frac{{\mathcal F}(L) 
- {\mathcal F}(0)}{\alpha_0 - \alpha_L} \,,\\
\label{eqn:dsol}
d &=& \frac{k}{H^2} \frac{\alpha_0 {\mathcal F}(L) - \alpha_L {\mathcal F}(0)}
{\alpha_0 - \alpha_L} \,,
\eea
with
\be
\alpha_i = \frac{2(1-\theta_i T_i) 
- \theta_i k \lambda_i (1-\theta_i T_i)^2}{1 + k\lambda_i T_i}\,, 
\ee
where we have borrowed a notation from \cite{clp}:
$\theta_0 = -1$, $\theta_L = +1$. The above formulas show a residual
gauge freedom parametrized by a single real function $F(y)$,
such that ${\mathcal F}'(y) = F(y)$ and 
${\mathfrak F}'(y) = {\mathcal F}/a^2(y)$.

\end{enumerate}

Using the background solutions, we can expand the action (\ref{eqn:ouraction}) 
up to the second order in an arbitrary metric fluctuation $h^{MN}$. 
The bulk part becomes
\bea\label{eqn:bulksecorderact1}
&& 2\int d^4 x \int_0^L d y \sqrt{-g} \Big(-8k^2 
+ k^2 h^2 + \frac{1}{2} R_{MPNQ} h^{MN} h^{PQ} \nn\\
&&\qquad -\frac{1}{4} \nabla_P h^{MN} \nabla^P h_{MN} 
+ \frac{1}{4} \nabla_M h \nabla^M h 
- \frac{1}{2} \nabla_M h \nabla_N h^{MN} 
+ \frac{1}{2} \nabla_M h^{MN} \nabla_P h^P_N \nn\\ 
&&\qquad + \nabla_M (\nabla_N h^{MN} - \nabla^M h 
- \frac{3}{2} h^{MN} \nabla^P h_{PN} 
- \frac{1}{2} h^{PN} \nabla_P h^M_N \nn\\
&&\qquad\qquad + h^{MN} \nabla_N h + \frac{1}{2} h \nabla_N h^{MN} 
+ h^{PN} \nabla^M h_{PN} - \frac{1}{2} h \nabla^M h )  \Big) \,,
\eea
the brane part is, with 4d total divergence terms dropped, 
\bea\label{eqn:branesecorderact1}
&&\sum_i \int d^4 x \Big[\sqrt{-g} 
\Big(-\frac{12\lambda_i H^2}{a^2} - U_i 
- \Big(\frac{3\lambda_i H^2}{a^2} + \frac{U_i}{2}\Big) \tilde h \nn\\
&&\qquad + \lambda_i (-\frac{1}{4} \tn_\rho h^{\mu\nu} 
\tn^\rho h_{\mu\nu} 
+ \frac{1}{4} \tn_\mu \tilde h \tn^\mu \tilde h 
- \frac{1}{2} \tn_\mu \tilde h \tn_\nu h^{\mu\nu} 
+ \frac{1}{2} \tn_\mu h^{\mu\nu} \tn_\rho h^\rho_\nu) \nn\\
&&\qquad + \Big(\frac{2\lambda_i H^2}{a^2} 
+ \frac{U_i}{4}\Big) h^{\mu\nu} h_{\mu\nu} 
- \Big(\frac{\lambda_i H^2}{2a^2} 
+ \frac{U_i}{8} \Big) \tilde h^2 \Big)\Big]_{y=y_i} \,,
\eea
and the extrinsic curvature part turns into
\bea\label{eqn:ksecorderact1}
&&2 \int d^4 x \Big[\sqrt{-g} \Big(\frac{8a'}{a} 
+ \tilde h' + \frac{4a'}{a} \tilde h - \frac{4a'}{a} h_{44} \nn\\
&&\qquad + \frac{1}{4} \tilde h^2{}' + \frac{a'}{a} \tilde h^2 
- h^{\mu\nu} h_{\mu\nu}' - \frac{2a'}{a} \tilde h h_{44} 
- \frac{1}{2} \tilde h{}' h_{44} 
+ \frac{3a'}{a} h_{44}^2 \Big) \Big]_{y=0}^{y=L}\,.
\eea
A tilde indicates that the corresponding entity is a 4d quantity 
constructed with $g_{\mu\nu}$. 
Note that there are 5d-total derivative terms in the bulk part. 
Due to the finiteness of the 5th dimension, they do not vanish identically 
but make contribution to the brane-boundary part of the action. 

Expanding further the bulk part, with the help of
\bea\label{eqn:cdc1}
\nabla_\mu T^\nu = \tn_\mu T^\nu 
+ \frac{a'}{a} \delta^\nu_\mu T^4\,, \quad
\nabla_\mu T^4 = \tn_\mu T^4 - \frac{a'}{a} T_\mu\,,
\eea
and imposing the partial gauge choice $h_{\mu4} = 0$, 
(\ref{eqn:bulksecorderact1})+(\ref{eqn:branesecorderact1})
+(\ref{eqn:ksecorderact1}) becomes 
\bea
\frac{S}{2M^3} &=& 2\int d^4 x \int_0^L d y \sqrt{-g} \Big(-8k^2 
- \frac{1}{2} h^{\mu\nu} \tn_\mu \tn_\rho h^\rho_\nu 
- \frac{1}{4} \tilde h \tn^2 \tilde h \nn\\
&&\quad + \frac{1}{4} h^{\mu\nu} \tn^2 h_{\mu\nu} 
+ \frac{1}{2} h^{\mu\nu} \tn_\mu\tn_\nu \tilde h 
- \frac{1}{2} \tilde h \tn^2 h_{44} 
+ \frac{1}{2} h_{44} \tn_\mu\tn_\nu h^{\mu\nu} \nn\\
&&\quad - \frac{1}{4} h^{\mu\nu}{}' h_{\mu\nu}{}' 
+ \frac{H^2 - 2a'{}^2}{2a^2} h^{\mu\nu} h_{\mu\nu} 
+ \frac{k^2 a^2 + a'{}^2}{2a^2} \tilde h^2 + \frac{\tilde h'{}^2}{4} 
+ \frac{a'}{4a}\tilde h^2{}' \nn\\
&&\quad + \frac{a'}{a} h_{44}^2{}' + \frac{k^2 a^2 + 6a'{}^2}{a^2} h_{44}^2 
- \frac{a'}{2a} \tilde h h_{44}{}' 
- \frac{2a'}{a} \tilde h' h_{44} + \frac{k^2 a^2 - 3 a'{}^2}{a^2} 
\tilde h h_{44} \Big) \nn\\
&&+ \sum_i \int d^4 x \Big[\sqrt{-g} \Big(-\frac{4\lambda_i H^2}{a^2} 
+ \frac{U_i}{3} \nn\\
&&\quad + \lambda_i (- \frac{1}{2} h^{\mu\nu} \tn_\mu 
\tn_\rho h^\rho_\nu 
- \frac{1}{4} \tilde h \tn^2 \tilde h + \frac{1}{4} h^{\mu\nu}
 \tn^2 h_{\mu\nu} 
+ \frac{1}{2} h^{\mu\nu} \tn_\mu\tn_\nu \tilde h) \nn\\
&&\quad + \frac{\lambda_i H^2}{2a^2} h^{\mu\nu} h_{\mu\nu} 
- \frac{U_i}{24} \tilde h^2 
- 2k T_i h_{44}^2 + k T_i \tilde h h_{44} \Big) \Big]_{y=y_i}\,.
\eea
Note that all the terms linear in $h$ get cancelled, as they should be. 

To get a simpler form, we remove $y$-derivatives on the 
fields whenever it's possible. For example,
\bea\label{eqn:ex1}
\sqrt{-g} \,\frac{a'}{4a} 
\tilde h^2{}' &=& \Big(\frac{a^3 a'}{4(1 - \frac{H^2 x^2}{4})^4} 
\tilde h^2 \Big)' 
- \frac{\tilde h^2}{4(1 - \frac{H^2 x^2}{4})^4} (3a^2 {a'}^2 + a^3 a'') \nn \\
&=& \Big(\sqrt{-g} \,\frac{a'}{4a} \tilde h^2 \Big)' 
- \sqrt{-g} \, \frac{3 {a'}^2 + k^2 a^2}{4a^2}  \tilde h^2\,,
\eea
where the first term contributes to the brane-boundary part of the action. 
This way, we eliminate $\tilde h^2{}'$, $h_{44}^2{}'$ 
and $\tilde h h_{44}'$--terms 
from the bulk part of the action, to get
\bea\label{eqn:secorderact2}
\frac{S}{2M^3} &=& 2\int d^4 x \int_0^L d y \sqrt{-g} \Big(-8k^2 
+ \frac{H^2 - 2a'{}^2}{2a^2} h^{\mu\nu} h_{\mu\nu} 
+ \frac{H^2}{4a^2} \tilde h^2 \nn\\
&&\quad - \frac{1}{2} h^{\mu\nu} \tn_\mu \tn_\rho h^\rho_\nu 
- \frac{1}{4} \tilde h \tn^2 \tilde h 
+ \frac{1}{4} h^{\mu\nu} \tn^2 h_{\mu\nu} 
+ \frac{1}{2} h^{\mu\nu} \tn_\mu\tn_\nu \tilde h \nn\\
&&\quad- \frac{1}{4} h^{\mu\nu}{}' h_{\mu\nu}{}' + \frac{\tilde h'{}^2}{4} 
- \frac{1}{2} \tilde h \tn^2 h_{44} 
+ \frac{1}{2} h_{44} \tn_\mu\tn_\nu h^{\mu\nu} \nn\\
&&\quad + \frac{3a'{}^2}{a^2} h_{44}^2  
- \frac{3a'}{2a} \tilde h' h_{44} + \frac{3H^2}{2a^2} 
\tilde h h_{44} \Big) \nn\\
&&+ \sum_i \int d^4 x \Big[\sqrt{-g} 
\Big(-\frac{4\lambda_i H^2}{a^2} + \frac{U_i}{3} 
+ \frac{\lambda_i H^2}{2a^2} h^{\mu\nu} h_{\mu\nu} 
+ \frac{\lambda_i H^2}{4a^2} \tilde h^2\nn\\
&&\qquad + \lambda_i (- \frac{1}{2} h^{\mu\nu} 
\tn_\mu \tn_\rho h^\rho_\nu 
- \frac{1}{4} \tilde h \tn^2 \tilde h \nn\\
&&\qquad\quad\quad + \frac{1}{4} h^{\mu\nu} \tn^2 h_{\mu\nu} 
+ \frac{1}{2} h^{\mu\nu} \tn_\mu\tn_\nu 
\tilde h) \Big) \Big]_{y=y_i}\,.
\eea
Plugging (\ref{eqn:massiveh}) and (\ref{eqn:masslessh}), we finally obtain
\be\label{eqn:secorderactfin}
\frac{S}{2M^3} 
= \frac{S}{2M^3}\Big|_{\Lambda} + \frac{S}{2M^3}\Big|_{\rm massive} 
+ \frac{S}{2M^3}\Big|_{\rm massless} \,, 
\ee
where
\bea\label{eqn:constaction}
\hspace{-15pt} \frac{S}{2M^3}\Big|_{\Lambda} 
&=& -16k^2 \int d^4 x \sqrt{-\hat g} \int_0^L d y \,a^4 
+ \sum_i \int d^4 x \sqrt{-\hat g} 
\Big[a^4 \Big(-\frac{4\lambda_i H^2}{a^2} 
+ \frac{U_i}{3}\Big)\Big]_{y=y_i} \nn\\
\hspace{-15pt} &=& -6H^2 {\mathcal C}_g^{(0)} \,,
\eea
\bea\label{eqn:massiveaction}
&&\frac{S}{2M^3}\Big|_{\rm massive} 
= 2\sum_{q > 0} \int d^4 x \int_0^L d y \sqrt{-g} 
\Big( \frac{1}{4} b^{(q)\mu\nu} \tn^2 b^{(q)}_{\mu\nu} 
- \frac{1}{4} b^{(q)\mu\nu}{}' b^{(q)}_{\mu\nu}{}' 
+ \frac{H^2-2a'{}^2}{2a^2} b^{(q)\mu\nu} b^{(q)}_{\mu\nu}\Big) \nn\\
&&\hspace{70pt} + \sum_{q > 0} \sum_i \int d^4 x \Big[\sqrt{-g} 
\Big(\frac{\lambda_i}{4} b^{(q)\mu\nu} \tn^2 b^{(q)}_{\mu\nu} 
+ \frac{\lambda_i H^2}{2a^2} b^{(q)\mu\nu} 
b^{(q)}_{\mu\nu}\Big)\Big]_{y=y_i} \nn\\
&&= \sum_{q>0} {\mathcal C}^{(q)}_g \int d^4 x \sqrt{-\hat g} \Big(
\frac{1}{4} B^{(q)\hat\mu\hat\nu} \hat\nabla^2 B^{(q)}_{\mu\nu} 
+ \frac{H^2}{2} B^{(q)\hat\mu\hat\nu} B^{(q)}_{\mu\nu} 
- \frac{m^{(q)}{}^2}{4} B^{(q)\hat\mu\hat\nu} B^{(q)}_{\mu\nu} \Big)\,,
\eea
and
\bea\label{eqn:masslessaction}
&&\frac{S}{2M^3}\Big|_{\rm massless} 
= 2\int d^4 x \int_0^L d y \sqrt{-g} 
\Big( \frac{1}{4} \beta^{\mu\nu} \tn^2 \beta_{\mu\nu} 
- \frac{1}{4} \beta^{\mu\nu}{}' \beta_{\mu\nu}{}' 
+ \frac{H^2-2a'{}^2}{2a^2} \beta^{\mu\nu} \beta_{\mu\nu}\Big) \nn\\
&&\hspace{70pt} + \sum_i \int d^4 x \Big[\sqrt{-g} 
\Big(\frac{\lambda_i}{4} \beta^{\mu\nu} \tn^2 \beta_{\mu\nu} 
+ \frac{\lambda_i H^2}{2a^2} \beta^{\mu\nu} 
\beta_{\mu\nu}\Big)\Big]_{y=y_i} \nn\\
&&\hspace{70pt} + 2\int d^4 x \int_0^L d y \sqrt{-g} 
\Big(-\frac{3}{2} {\mathcal Y}_2^2 
- \frac{3}{4} H^2 a^2 {\mathcal Y}_1'{}^2 
- \frac{3}{2} F {\mathcal Y}_2 \Big) \psi {\mathcal D}_4 \psi \nn\\
&&\hspace{70pt} + \sum_i \int d^4 x \Big[\sqrt{-g} 
\Big(- \frac{3\lambda_i}{2} {\mathcal Y}_2^2 \Big) 
\psi {\mathcal D}_4 \psi \Big]_{y=y_i} \nn\\
&&= {\mathcal C}^{(0)}_g \int d^4 x \sqrt{-\hat g} \,\Big(
\frac{1}{4} B^{\hat\mu\hat\nu} \hat\nabla^2 B_{\mu\nu} 
+ \frac{H^2}{2} B^{\hat\mu\hat\nu} B_{\mu\nu} \Big) 
+ {\mathcal C}_\psi \int d^4 x \sqrt{-\hat g} \,\psi 
\hat{\mathcal D}_4 \psi\,,
\eea
with
\bea\label{eqn:zerogravcoeffi}
{\mathcal C}^{(0)}_g &=& \frac{1-T_0^2}{k}\sum_i \Big(\tanh^{-1} T_i 
+ \frac{T_i + k\lambda_i}{1-T_i^2}\Big) \,, \\
\label{eqn:cgq}
{\mathcal C}^{(q)}_g &=& \frac{k}{H^2} \Big( 2\int_{-T_0}^{T_L} 
{\mathcal Y}^{(q)}{}^2 dz 
+ \sum_i [v_i (1-z^2) {\mathcal Y}^{(q)}{}^2]_{y=y_i} \Big)\,, \\
\label{eqn:mq}
m^{(q)}{}^2 &=& H^2 \frac{\int_{-T_0}^{T_L} 
\frac{\{\partial_z ((1-z^2){\mathcal Y}^{(q)})\}^2}{1-z^2} dz}
{\int_{-T_0}^{T_L} {\mathcal Y}^{(q)}{}^2 dz 
+ \sum_i \frac{v_i}{2} [(1-z^2){\mathcal Y}^{(q)}{}^2]_{y=y_i}}\,, \\
\label{eqn:genradioncoeffi}
{\mathcal C}_\psi &=& - \int_0^L d y \,a^2 \Big(3 {\mathcal Y}_2^2 
+ \frac{3}{2} H^2 a^2 {\mathcal Y}_1'{}^2 + 3 F {\mathcal Y}_2 \Big) 
- \sum_i \Big[\frac{3\lambda_i}{2} a^2 {\mathcal Y}_2^2 \Big]_{y=y_i} \,.
\eea
A hatted entity is defined using the metric without a warp factor.
If we include a source term in the action, its coupling to 
a specific graviton mode will show up like
\be
\hspace{-20pt} S^{(q)} &=& 2M^3 {\mathcal C}^{(q)}_g 
\int d^4 x \sqrt{-\hat g}\; \frac{1}{4} B^{(q)\hat\mu\hat\nu} 
\hat\nabla^2 B^{(q)}_{\mu\nu} 
+ \cdots + \int d^4 x dy \sqrt{-g}\; h^{(q)}_{\mu\nu} T^{\mu\nu} \nn\\
\hspace{-20pt} &=& \int d^4 x \sqrt{-\hat g} \; \frac{1}{4} 
\bar B^{(q)\hat\mu\hat\nu} \hat\nabla^2 \bar B^{(q)}_{\mu\nu} 
+ \cdots + \int d^4 x dy \sqrt{-g}\; 
\frac{{\mathcal Y}^{(q)}(y)}{\sqrt{2M^3 {\mathcal C}^{(q)}_g}} 
\bar B^{(q)}_{\mu\nu} T^{\mu\nu} \,, 
\ee
and then we can read off the gravitational coupling constant. 
In particular, on the 0-brane,
\be\label{eqn:gcq0}
\frac{1}{M_4^{(q)}} = \frac{{\mathcal Y}^{(q)}(0)}{\sqrt{2M^3 
{\mathcal C}^{(q)}_g}}\,.
\ee

\section{Avoiding ghosts and tachyons}

\subsection{ghostbusting}

It has been known for some time that brane setups of the type that we
are considering are sometimes afflicted with unphysical features.
Kinetic ghosts, by which we mean wrong-sign kinetic terms for physical
modes in the 4d effective action, are indicative of an instability,
similar to the case of their cousin the tachyon. Regions of our input
parameter space which produce kinetic ghosts are certainly to be avoided
if we are interested in static brane configurations. A kinetic ghost radion
can occur for setups with \eg, $v_0=v_L=0$ and $w_0 + w_L < 0$. 
Intuitively we also expect
a kinetic ghost graviton to occur in cases where $v_0$ and $v_L$ become
too negative, {\it i.e.} we have too much wrong-sign localized curvature.
As we will see, the full story is quite complicated.

The boundary between a region of the input parameter space which
has a kinetic ghost, and a region which does not, defines a class of
models where the coefficient of the kinetic term of a physical mode
is vanishing. After a canonical rescaling of the field, this implies
strong coupling once we go beyond the linearized theory. Such regions
of strong coupling are to be avoided if we want the 4d effective low
energy theory to be valid up to energy scales approaching $k$.

In this section we will map out the input parameter space and
identify the region which avoids both kinetic ghosts and strong coupling.

\begin{figure}
  \begin{center}
    \resizebox{11cm}{!}{\includegraphics{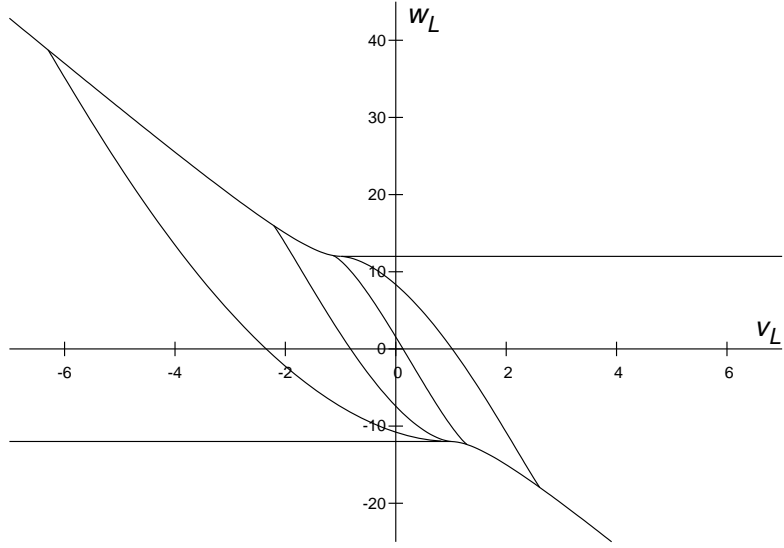}}
  \end{center}
  \caption{The $\bar{\mathcal C}^{(0)}_g=0$ on $(v_L,w_L)$-plane when
$v_0=0.3$ and $w_0=10,5,-5,-10$ for the lines from left to right respectively.}
  \label{grav2}
\end{figure}

\begin{figure}
  \begin{center}
    \resizebox{11cm}{!}{\includegraphics{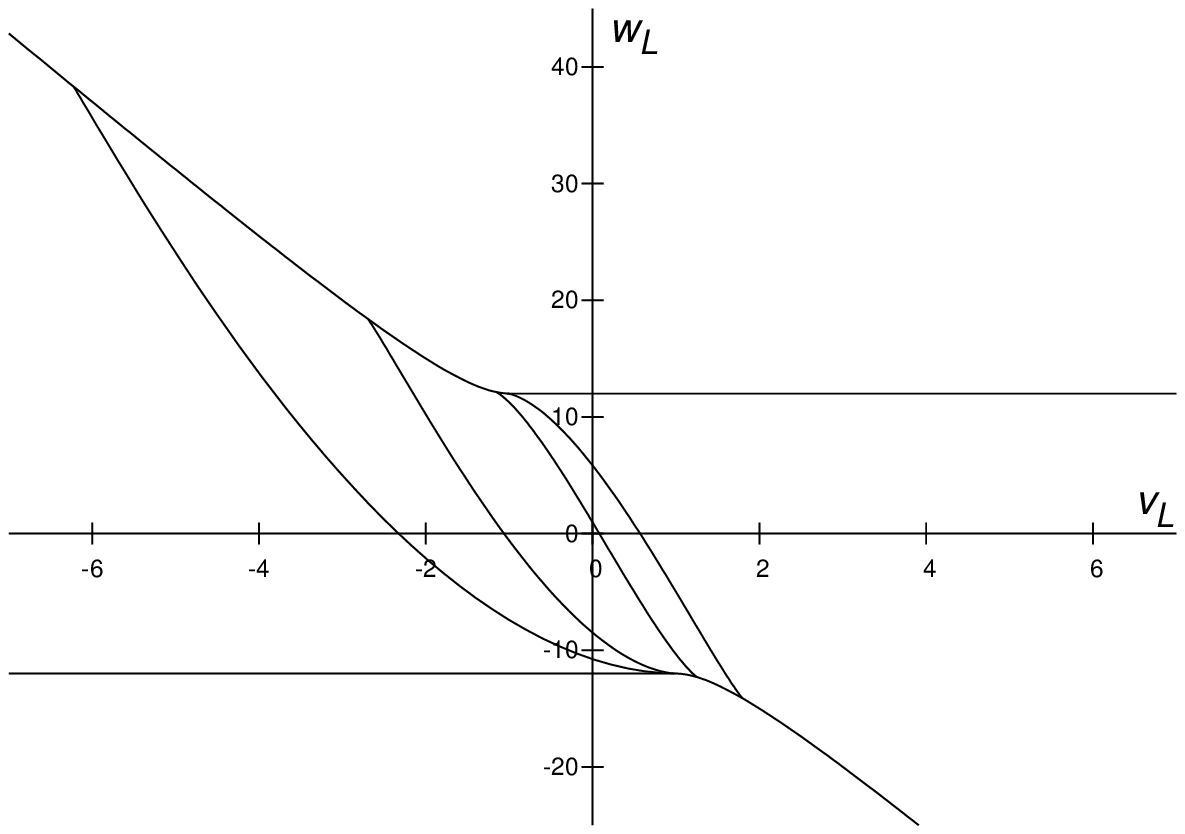}}
  \end{center}
  \caption{The $\bar{\mathcal C}^{(0)}_g=0$ on $(v_L,w_L)$-plane when
$w_0=5$ and $v_0=1.5,0.5,-0.5,-1$
for the lines from left to right respectively.}
  \label{grav3}
\end{figure}

We have already noted that the massless mode of the graviton 
may be a ghost.
When we choose $(v_0, w_0)$ such that $T_0 > 0$, 
Figure \ref{grav2} and \ref{grav3} 
show how $\bar{\mathcal C}^{(0)}_g 
\equiv k {\mathcal C}^{(0)}_g / (1-T_0^2)$ varies 
as a function of $(v_L, w_L)$. $\bar{\mathcal C}^{(0)}_g$ is zero along each 
line shown in Figure \ref{grav2} and \ref{grav3}, 
positive {\it above} it and negative {\it below} it. 
As $T_0$ approaches $1$ either by $v_0 \to +\infty$ or by $w_0 \to +12$, 
$\bar{\mathcal C}^{(0)}_g = 0$ line moves to the left. 

\begin{figure}
  \begin{center}
    \resizebox{11cm}{!}{\includegraphics{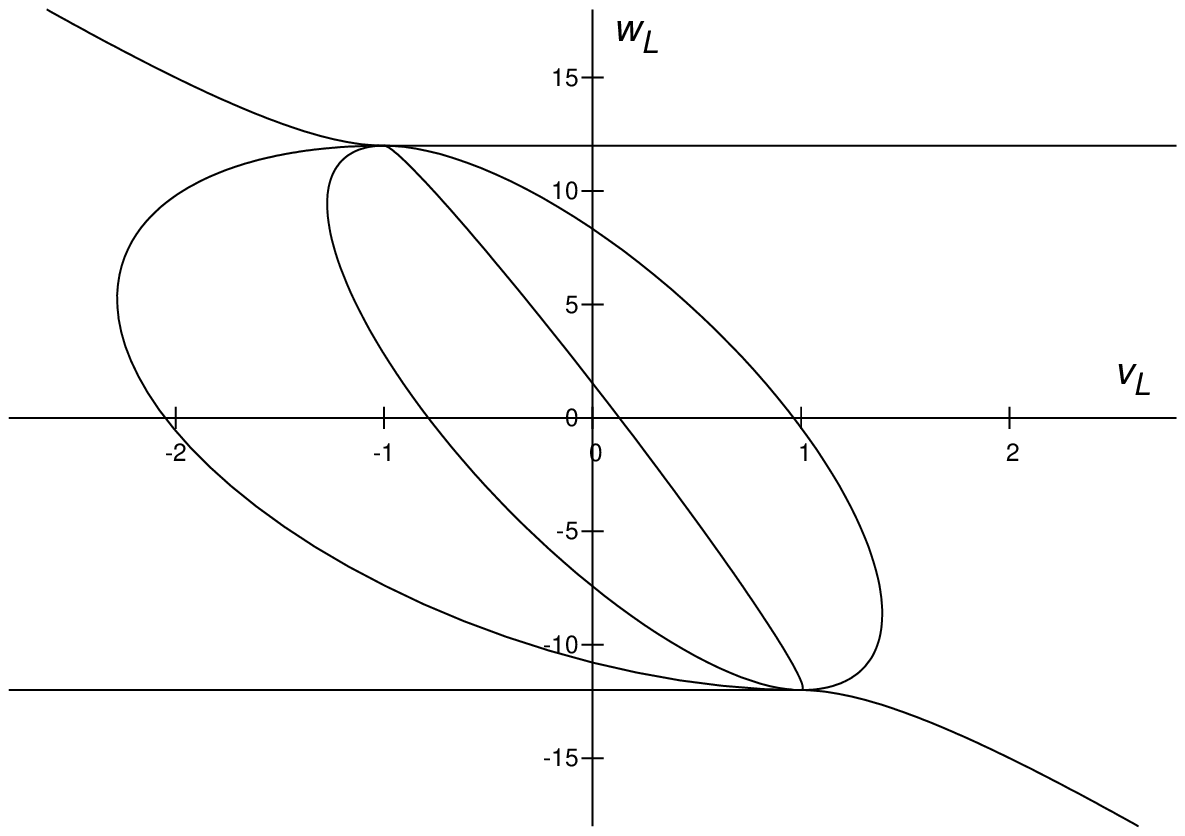}}
  \end{center}
  \caption{The $\bar{\mathcal C}_\psi=0$ on $(v_L,w_L)$-plane when
$v_0=0.3$ and $w_0=10,5,-5,-10$ for the lines from left to right respectively.}
  \label{rad2}
\end{figure}

\begin{figure}
  \begin{center}
    \resizebox{11cm}{!}{\includegraphics{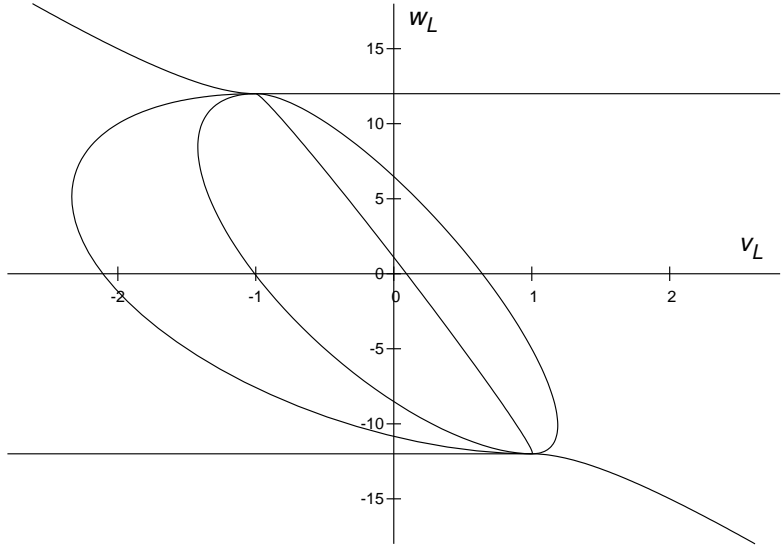}}
  \end{center}
  \caption{The $\bar{\mathcal C}_\psi=0$ on $(v_L,w_L)$-plane when
$w_0=5$ and $v_0=1.5,0.5,-0.5,-1$
for the lines from left to right respectively.}
  \label{rad3}
\end{figure}

We can play a similar game with the coefficient, ${\mathcal C}_\psi$, 
of the radion. Equation
(\ref{eqn:genradioncoeffi}) holds irrespective of gauge choice of $F(y)$. 
Then, in a generic case where $\alpha_0 - \alpha_L \neq 0$, for simplicity 
we can choose
\bea\label{eqn:fchoice}
{\mathcal F}(y) = \chi \Big( \frac{y}{L} (\alpha_L - \alpha_0) 
+ \alpha_0 \Big) \,,
\eea
where $\chi$ is a remaining real gauge parameter
that we leave arbitrary as a check of general covariance
for physical results. With this choice we get
\be
c = - \frac{1}{H^2} \,,\;\; d = 0\,.
\ee
Then, (\ref{eqn:genradioncoeffi}) becomes
\be\label{eqn:radioncoeffi}
{\mathcal C}_\psi &=& \frac{3 \chi^2 H^2}{k} \sum_i \Big[ 
-\frac{1}{2} (\theta_i + v_i z) \alpha_i 
\Big(\frac{z}{1-z^2} \alpha_i + 2 \frac{1-z}{1+z} \Big) 
+ \Big(\theta_i - \frac{1-z}{2} v_i \Big) \frac{(1-z)^2}{1+z} 
\Big]_{z=\theta_i T_i} \nn\\
&=& \frac{3 \chi^2 H^2}{2 k} \sum_i \frac{(-\theta_i + T_i)(2 
- \theta_i v_i + v_i T_i)}{1 + v_i T_i}\,.
\ee
Figures \ref{rad2} and \ref{rad3} show that, if we choose $(v_0, w_0)$ 
such that $T_0 > 0$, 
then $\bar {\mathcal C}_\psi \equiv 2k {\mathcal C}_\psi / 3 \chi^2 H^2$ 
is positive above each line. 
Also, the closer $T_0$ is to $+1$, the more convex to the 
left $\bar {\mathcal C}_\psi = 0$ line gets.

\subsection{tachyons}
Equation (\ref{eqn:deteq}) determines the KK mass spectrum of the graviton.
For some range of values of the input parameters, 
it can have a zero at negative $q$, \textit{i.e.} negative mass-squared.
For example, if we choose $v_0=5, w_0=3, v_L=-7, w_L=9$, 
then (\ref{eqn:deteq}) has a solution $q=-0.94$. 

We will refer to such solutions as tachyons. The tachyons with
$-9/4 < q < 0$ obey the 
Breitenlohner-Freedman bound \cite{Breitenlohner:1982jf},
and as expected we will find (see section 5.1)
that the Euclidean Green's function for
such modes has the same exponentially damped asymptotics as
for ordinary massive modes. The tachyons with $q < -9/4$
have oscillatory asymptotics, similar to tachyons in flat space. 
It turns out that some of the tachyonic solutions also have wrong-sign
kinetic terms.

We can expose the tachyonic solutions by analyzing the quantity
\be
{\mathfrak D} \equiv \frac{a_0 b_L - b_0 a_L}{q (1-T_0^2)(1-T_L^2)}\,.
\ee
Note that ${\mathfrak D}$ takes some finite value at $q=0$,
proportional to the residue of the pole from the massless graviton mode.
Now we can compare the two quantities
\bea
{\mathfrak D} |_{q=0} \,,\qquad {\mathfrak D} |_{q\to-\infty} \,.
\eea
Suppose we find values for the input parameters such that there
are no tachyons in the KK graviton spectrum. Then it must be
the case that
\be\label{eqn:notachyon}
{\mathfrak D} |_{q=0} \times {\mathfrak D} |_{q\to-\infty} > 0 \; .
\ee
As we vary the input parameters, we may cross the hypersurface in the
parameter space defined by 
${\mathfrak D} |_{q=0} \times {\mathfrak D} |_{q\to-\infty} = 0$.
Just across this boundary are models which each contain a single tachyon.
To find this boundary between models with
no tachyon and models with a single tachyon,
we need explicit expressions for ${\mathfrak D} |_{q=0}$ and
${\mathfrak D} |_{q\to-\infty}$.
 
${\mathfrak D} |_{q=0}$ can be worked out straightforwardly:
\bea\label{eqn:detat0}
{\mathfrak D} |_{q=0} = -2 \sum_i
\Big( \frac{1}{2}\log \frac{1+T_i}{1-T_i} + \frac{T_i+v_i}{1-T_i^2} \Big) 
= -2 \bar{\mathcal C}_g^{(0)} \,,
\eea
where we used $\tanh^{-1}x=\frac{1}{2}\log\frac{1+x}{1-x}$.

Before trying to evaluate ${\mathfrak D} |_{q\to-\infty}$, 
note that our solutions 
of the equations of motion should be real. 
While $P_{(-1+\sqrt{9+4q})/2}^{-2}(z)$ 
is always real, $Q_{(-1+\sqrt{9+4q})/2}^2(z)$ becomes 
a complex valued function for $q< -9/4$.
Thus for $q > -9/4$ we still write (\ref{eqn:massiveysol}) in the
form
\be\label{eqn:realmassiveysol}
{\mathcal Y}^{(q)}(y) = A P_{(-1+\sqrt{9+ 4q})/2}^{-2}(z) 
+ B Q_{(-1+\sqrt{9+ 4q})/2}^2(z) 
\,,
\ee
but for $q < -9/4$ we use relation 8.843 of
\cite{gradrhyz} and relations
3.6.1(4)-(5), 3.3.1(7) of \cite{Bateman} to write
the real solution
\be\label{eqn:-qmassiveysol}
{\mathcal Y}^{(q)}(y) = A P_{ip-1/2}^{-2}(z) 
+ B \frac{\pi}{2\cosh \pi p}(p^2+\frac{9}{4})(p^2+\frac{1}{4}) 
P_{ip-1/2}^{-2}(-z) \,,
\ee
where we have defined $2ip=\sqrt{9+ 4q}$.
The two expressions match at $p=0$, $q=-9/4$.
 
Plugging (\ref{eqn:-qmassiveysol})
into the brane-boundary equations of motion, 
we obtain $a_0$, $b_0$, $a_L$, $b_L$ which are real 
for any large and negative $q$, 
\ie, large and positive $p$:
\bea
a_0 &=& \sqrt{1-T_0^2}\;
\Big\{-v_0 (p^2+\frac{9}{4})\sqrt{1-T_0^2}\, P_{ip-1/2}^{-2}(-T_0) 
- 2 P_{ip-1/2}^{-1}(-T_0)\Big\} \,,\nn\\
b_0 &=& \sqrt{1-T_0^2}\;
\Big\{-v_0 (p^2+\frac{9}{4})\sqrt{1-T_0^2}\, P_{ip-1/2}^{-2}(T_0) 
+ 2 P_{ip-1/2}^{-1}(T_0)\Big\} \,,\nn\\
a_L &=& \sqrt{1-T_L^2}\;
\Big\{-v_L (p^2+\frac{9}{4})\sqrt{1-T_L^2}\, P_{ip-1/2}^{-2}(T_L) 
+ 2 P_{ip-1/2}^{-1}(T_L)\Big\} \,,\\ 
b_L &=& \sqrt{1-T_L^2}\;
\Big\{-v_L (p^2+\frac{9}{4})\sqrt{1-T_L^2}\, P_{ip-1/2}^{-2}(-T_L) 
- 2 P_{ip-1/2}^{-1}(-T_L)\Big\} \,. \nn
\eea
Using relations 3.2 (14), 3.4 (1), 2.3.2 (17)
from \cite{Bateman}, we can 
derive the asymptotic behavior of the conical
function $P_{ip-1/2}^{-m}(\cos \theta)$ for large $p$:
\be
P_{ip-1/2}^{-m}(\cos \theta) 
= \frac{1}{\sqrt{2\pi\sin\theta}} \frac{e^{p\theta}}{p^m\sqrt{p}} 
(1+{\mathcal O}(p^{-1}))\,,\qquad 0<\theta<\pi\,,
\ee
and thus we get
\bea
{\mathfrak D} |_{q\to-\infty} \approx -\frac{1}{p^2} \frac{v_0 v_L}
{2\pi p \sqrt{\sin\theta_0\sin\theta_L}}
\Big( e^{p(2\pi-\theta_0-\theta_L)} - e^{p(\theta_0+\theta_L)} \Big) \,, 
\eea
where $\cos \theta_i = T_i$. 

Now we are ready to look at (\ref{eqn:notachyon}), which implies
\bea\label{eqn:wrong}
{\mathfrak D} |_{q=0} > 0 \quad {\rm and} 
\quad {\mathfrak D} |_{q\to-\infty} > 0\,,
\eea
or
\bea\label{eqn:right}
{\mathfrak D} |_{q=0} < 0 
\quad {\rm and} \quad {\mathfrak D} |_{q\to-\infty} < 0\,.
\eea
Note that the sign of ${\mathfrak D} |_{q\to-\infty}$
is the same as the sign of
\be\label{eqn:notachyon2}
v_0 v_L (\cos^{-1}T_0 + \cos^{-1}T_L - \pi) \,.
\ee

\begin{figure}
  \begin{center}
    \resizebox{13cm}{!}{\includegraphics{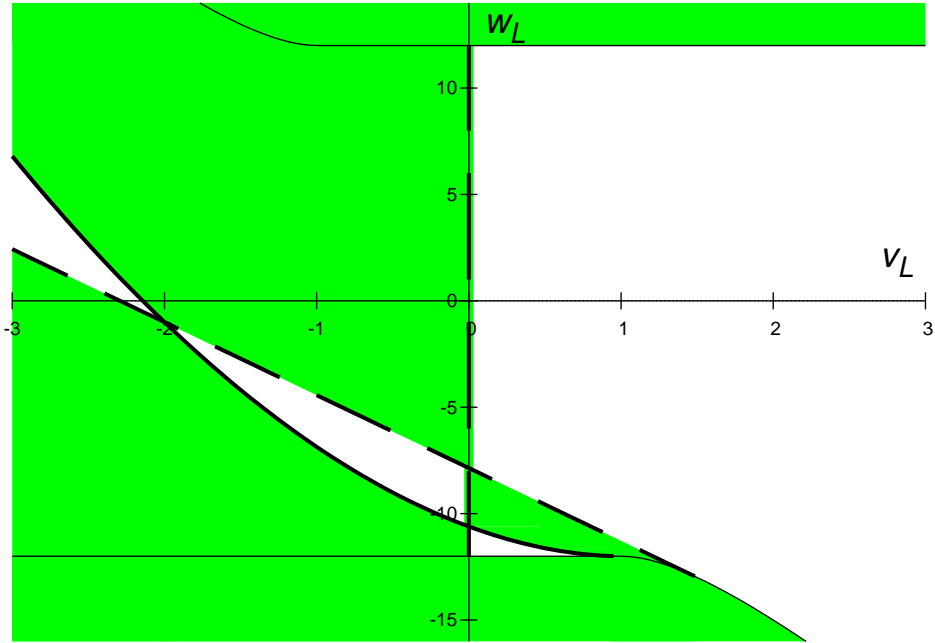}}
  \end{center}
  \caption{${\mathfrak D} |_{q=0} \times {\mathfrak D} |_{q\to-\infty} > 0$ 
  on $(v_L,w_L)$-plane when $v_0=2$ and $w_0=1$. 
The solid curve is where ${\mathfrak D} |_{q=0}=0$ 
and dashed straight lines are ${\mathfrak D} |_{q\to-\infty} = 0$.}
  \label{tachyon1}
\end{figure}

Figure \ref{tachyon1} shows the tachyon counting for models
in the $(v_L,w_L)$ plane
when $v_0 > 0$. The sign of
${\mathfrak D} |_{q=0}$ flips when we cross the line
$\bar{\mathcal C}_g^{(0)} = 0$, which 
is shown as the solid curve. The sign of
(\ref{eqn:notachyon2}) flips whenever we cross one of
the two dashed straight lines.
The vertical dashed line is just $v_L = 0$. The
slanted dashed line is
\be
w_L = -6(1-T_0^2)v_L -12 T_0 \; .
\ee

Thus the shaded region in Figure \ref{tachyon1}
gives models which have a single tachyon
(or are not $AdS_5/AdS_4$). The models in the four unshaded regions
have an even number of tachyons. Models in the large unshaded
region to the right of $v_L = 0$ are tachyon-free, as expected.

By further numerical analysis,
we find that models in the upper two remaining unshaded regions are
also tachyon-free, while models in the lower one have
two tachyons. 
Once we also require $\bar{\mathcal C}_g^{(0)} > 0$, only the
middle unshaded area remains.
This region is rather special. It contains no tachyons and
obeys $\bar{\mathcal C}_g^{(0)} > 0$. However in this region
one can see numerically that the massive KK graviton
modes are kinetic ghosts.

\begin{figure}
  \begin{center}
    \resizebox{13cm}{!}{\includegraphics{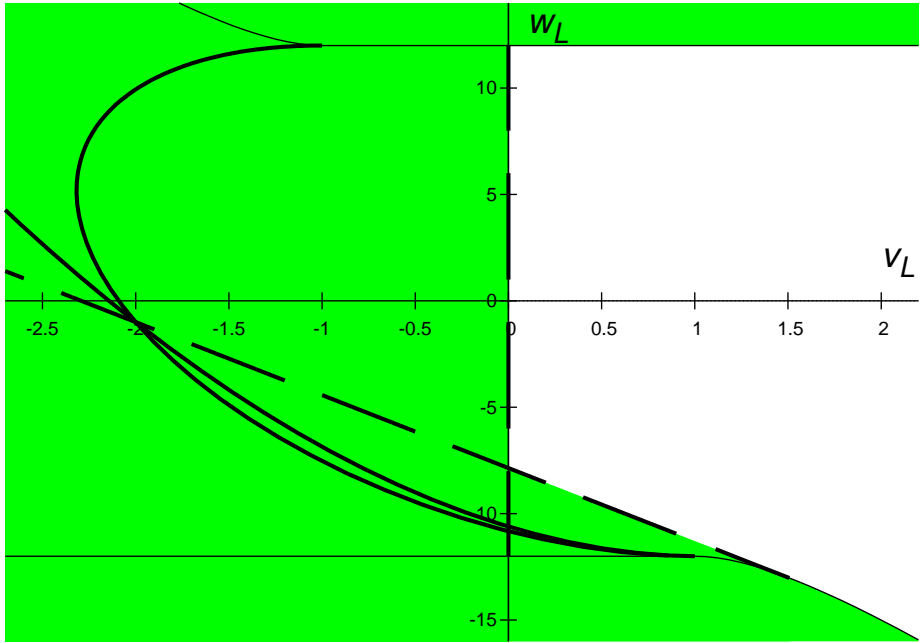}}
  \end{center}
  \caption{Allowed region (unshaded)
on $(v_L,w_L)$-plane with $v_0=2$ and $w_0=1$.}
  \label{combi1}
\end{figure}

Combining all of the above results with the additional requirement
that $\bar {\mathcal C}_\psi > 0$, \textit{i.e.} no ghost radion,  
we get figures like Figure \ref{combi1} for $(v_0,w_0)$ chosen
such that $T_0 > 0$.
Models which are ghost-free and tachyon-free 
correspond to the unshaded region in the $(v_L, w_L)$ plane.
For $v_0 < 0$, a similar analysis finds no region
which is free of tachyons and ghosts.

These results
fit qualitatively with our physical intuition. Models
with negative $M_i$ are excluded. Models with
too much negative tension branes are also excluded.

A special case is that of an infinite extra dimension.
In our general framework we make models of this type by
sending $L \to \infty$. We do this by fixing $w_L = +12$ and
(for simplicity) $v_L = 0$.
In this limit 
the graviton zero mode is not normalizable and drops out of the theory, 
so there is no ${\mathcal C}_g^{(0)}$ to consider. 
As for ${\mathcal C}_\psi$, since $T_L = +1$, 
(\ref{eqn:radioncoeffi}) becomes 
\be
{\mathcal C}_\psi = \frac{3 \chi^2 H^2}{2 k} 
\frac{(1 + T_0)(2 + v_0 + v_0 T_0)}{1 + v_0 T_0}\,,
\ee
which is positive for any $v_0$ and $w_0 < 12$.
Our tachyon analysis reduces to examining
$a_0 |_{q=0}$ and $a_0 |_{q\to-\infty}$. Since
\be
a_0 |_{q=0} = -(1-T_0^2) < 0 \,, \nn
\ee
and
\be
a_0 |_{q\to-\infty} \approx -v_0 
\frac{1-T_0^2}{\sqrt{2\pi\sin\theta_0}}
\frac{e^{p(\pi-\theta_0)}}{\sqrt{p}} \,, 
\ee
only $v_0 > 0$ is allowed. 
So in this class of models the allowed
region is $v_0 > 0$ and $w_0 < 12$.

\section{Strong coupling}
For generic choices of the
four input parameters $v_0$, $w_0$, $v_L$ and $w_L$
we will not have a strong coupling problem, 
as long as we choose parameters away from 
the borderline between ghost and non-ghost regions,
and away from limits where $H\to 0$.

To determine the strong coupling scale, 
we need to calculate the cubic action for the radion. 
See Appendix \ref{appA} 
for the details of the calculation. 
The general cubic expansion of the radion part of (\ref{eqn:ouraction}) is
\bea\label{eqn:r3}
\frac{S}{2M^3}\Big|_{\psi^3} 
&=& \int d^4x \sqrt{-g} 
\Big( {\mathcal C}^{(3)}_1 \hn^2 \psi \{(\hn^2 \psi)^2 
- \hn_\mu \hn_\nu \psi \hn^\mu \hn^\nu \psi \} \nn\\
&&\quad + {\mathcal C}^{(3)}_2 \psi \{ (\hn^2 \psi)^2 
- \hn_\mu \hn_\nu \psi \hn^\mu \hn^\nu \psi \} 
+ {\mathcal C}^{(3)}_3 \psi^2 \hn^2 \psi 
+ {\mathcal C}^{(3)}_4 \psi^3 \Big) \,.
\eea
When it comes to the strong coupling, 
${\mathcal C}^{(3)}_1$ would be expected to give the
most stringent limit. However
with $F(y)$ determined by (\ref{eqn:fchoice}), 
direct calculation shows that
\bea
{\mathcal C}^{(3)}_1 &=& \sum_i 
\Big[\Big( \frac{\theta_i + v_i z}{2(1-z)} {\mathcal F}(y) 
- \chi (\theta_i - \frac{1-z}{2} v_i ) \Big) \nn\\
&& \qquad \Big( \frac{H^2}{4k^2} \frac{{\mathfrak F}(y)^2}{1+z} 
+ \frac{\chi}{2k} \frac{(1-z)^2}{1+z} {\mathfrak F}(y) 
+ \frac{\chi^2}{4H^2} \frac{(1-z)^4}{1+z} \Big)  \Big]_{y=y_i} \nn\\
&=& 0\,,
\eea
The next candidate, 
${\mathcal C}^{(3)}_2$, turns out to be
\bea\label{eqn:radioncubicc}
{\mathcal C}^{(3)}_2 &=& \sum_i \theta_i \Big[-\frac{{\mathcal F}(y)^3}{12} 
+ \frac{\chi H^2}{k^2} \frac{F {\mathfrak F}(y)}{1+z} \Big]_{y=y_i} 
+ \frac{2\chi H^2}{k} \int_0^L \frac{F {\mathfrak F}(y)}{1+z} dy \nn\\
&=& \chi^3 \frac{\alpha_0 - \alpha_L}{12} \Big\{ \alpha_0^2 
+ \alpha_0 \alpha_L + \alpha_L^2 
+ \frac{-12 \alpha_L T_0^{-1} -12 \alpha_0 T_L^{-1} + 6(\alpha_0 
- \alpha_L) \log4}{T_0^{-1} + T_L^{-1}} \nn\\
&&\qquad + \frac{6(\alpha_0 - \alpha_L)}{(T_0^{-1} + T_L^{-1})^2} 
\Big( {\rm Li}_2 (-\frac{1+T_0}{1-T_0}) 
- {\rm Li}_2 (-\frac{1-T_L}{1+T_L}) \Big) \Big\}\,,
\eea 
where $T_i^{-1} = \tanh^{-1} T_i$ and ${\rm Li}_n$ is a 
polylogarithm function. 
This does not vanish. Then from the full radion action 
\bea\label{eqn:dgpradaction}
S\,|_\psi &=& 2M^3 {\mathcal C}_\psi \int d^4 x \sqrt{-\hat g} 
\,\psi \hn^2 \psi \nn\\
&&+ 2M^3 {\mathcal C}^{(3)}_2 \int d^4 x \sqrt{-\hat g} 
\,\psi \{ (\hn^2 \psi)^2 
- \hn_\mu \hn_\nu \psi \hn^\mu \hn^\nu \psi \} + \cdots \nn\\
&=& \int d^4 x \sqrt{-\hat g} \,\bar\psi \hn^2 \bar\psi \nn\\
&&+ \frac{2M^3 {\mathcal C}^{(3)}_2}{(2M^3 {\mathcal C}_\psi)^{3/2}} 
\int d^4 x \sqrt{-\hat g} \,\bar\psi \{ (\hn^2 \bar\psi)^2 
- \hn_\mu \hn_\nu \bar\psi \hn^\mu \hn^\nu \bar\psi \} + \cdots\,,
\eea
we can determine the strong coupling scale:
\be\label{eqn:scs}
\Lambda_{\rm sc} = (2M^3)^{1/6} 
\frac{{\mathcal C}_\psi^{1/2}}{{\mathcal C}^{(3)}_2{}^{1/3}}\,.
\ee
As an example, let's consider a DGP-like limit ($v_L = 0$, 
$w_L = +12$, $v_0$ is large). Then (\ref{eqn:scs}) becomes
\be
\Lambda_{\rm sc} \sim \frac{M^2}{M_0} \sqrt{12-w_0} \,,
\ee
where our parameter $M_0$ is equivalent to the 
parameter $M_P$ in DGP.
This agrees well with \cite{LPR}.

In our general framework we can understand the robustness of the
strong coupling problem. As long as we restrict ourselves 
to models in the ghost+tachyon free region of the parameter space,
the coefficient $\bar{\mathcal C}_g^{(0)}$
of the graviton kinetic term never vanishes,
as can be seen from (\ref{eqn:zerogravcoeffi}).
From (\ref{eqn:scs}) we can see that
the radion becomes strongly coupled in any limit where
$H \to 0$. This includes the DGP-like limit just mentioned,
as well as the ``bigravity'' limit $T_0 \to +1$. Unfortunately
these are precisely the limits in which we find ultralight
graviton modes.

\section{Green's function analysis}
In \cite{clp}, the set of coupled equations of motion of the graviton and the 
radion was obtained in a straight gauge.
Once we eliminate the radion, we get four independent equations 
involving the graviton only. These are of the form:
\bea
&&{\mathcal D}^{(\rm i)}_{\mu\nu\rho\sigma} h^{\rho\sigma} = 0\,,\\
&&{\mathcal D}^{(\rm ii)}_{\mu\nu} h^{\mu\nu} = 0\,,\\
&&[{\mathcal D}^{(\rm bdy)}_{\mu\nu\rho\sigma} h^{\rho\sigma}]_{y=y_i} = 0\,,
\eea
where the full expressions are given in
(\ref{eqn:gravonlyeq1})-(\ref{eqn:gravonlybbeq}). These
imply the following Green's function equations in the straight gauge:
\bea\label{eqn:gfemunu}
&&{\mathcal D}^{(\rm i)}_{\mu\nu\rho\sigma} G^{\rho\sigma}_{;\mu'\nu'} = 
\frac{1}{M^3} O_{\mu\nu ;\mu'\nu'}(x,x') 
\frac{\delta^{(4)}(x-x')\delta(y-y')}{\sqrt{-g}} \,,\\
&&{\mathcal D}^{(\rm ii)}_{\mu\nu} G^{\mu\nu}_{;\mu'\nu'} = 0\,,\\
\label{eqn:gfebb}
&&[{\mathcal D}^{(\rm bdy)}_{\mu\nu\rho\sigma} 
G^{\rho\sigma}_{;\mu'\nu'}]_{y=y_i} = 0\,,
\eea
where the $AdS_4$ bitensor $O_{\mu\nu ;\mu'\nu'}(x,x')$ is given below.
Then, for any given source, we get the linearized solution for the graviton
from
\be
h_{\mu\nu}(x,y) 
= \int d^4x'dy \sqrt{-g} \, G_{\mu\nu;\mu'\nu'}(x,y;x',y') 
T^{\mu'\nu'}(x',y')\,.
\ee
In Appendix \ref{appB}, the Euclidean versions of
(\ref{eqn:gfemunu})-(\ref{eqn:gfebb})
are explicitly solved to obtain the Euclidean Green's function.  
After dropping 4d total derivatives which will vanish when contracted
with a conserved stress tensor, we can write the following
expression for the Euclidean Green's function on the 0-brane:
\bea\label{eqn:fullgreenftn}
&&\hspace*{-35pt}
G_{\mu\nu\,;\,\mu'\nu'}(x,x',y=y'=0) 
= \sum_j G_1(u;p_j) \bigg\{\,T^{(3)}_{\mu\nu\,;\,\mu'\nu'}  \nn\\
&&\hspace{45pt} 
-\frac{2}{3}\frac{H^4}{p_j^2-\frac{1}{4}}T^{(1)}_{\mu\nu\,;\,\mu'\nu'}
\Big( p_j^2-\frac{9}{4}+3(1+u)^2
+\frac{3(1+u)}{\frac{d}{du}{\rm ln}\,Q_{p_j-\frac{1}{2}}(1+u)} \Big)\bigg\}
\,,
\eea
where the sum is over the residues of poles from the individual
KK graviton modes. The masses $m_j$ of these modes are determined by
the solutions of (\ref{eqn:deteq}), and are expressed in terms
of the parameter $p_j$:
\bea
m_j^2 = H^2(p_j^2-\frac{9}{4})
\; .
\eea
The variable $u$ is related to the geodesic distance $\mu$
between the points $x$ and $x'$ in $AdS_4$:
$u = \cosh H\mu -1$. The function $G_1(u;p_j)$ is given by
\be\label{eqn:ourgone}
G_1(u;p_j) = -\frac{1}{4\pi^2 H^2 M_4^{(j)}{}^2}
\frac{d}{du} Q_{p_j-1/2}(1+u)\;,
\ee
where $Q_{p_j-1/2}$ is a Legendre function.
The $AdS_4$ tensor structure of the Green's function
(\ref{eqn:fullgreenftn}) is contained in the bitensors
$T^{(1)}$ and $T^{(3)}$, which are part of the complete
bitensor basis given in (\ref{eqn:Gdecomp1}):
\be
T^{(1)}_{\mu\nu\,;\,\mu'\nu'} &=& g_{\mu\nu}g_{\mu'\nu'} \; ,\nonumber\\
T^{(3)}_{\mu\nu\,;\,\mu'\nu'} &=&
\partial_{\mu}\partial_{\mu'}u\partial_{\nu}\partial_{\nu'}u +
\partial_{\mu}\partial_{\nu'}u\partial_{\nu}\partial_{\mu'}u \; .
\ee
The bitensor on the right hand side of (\ref{eqn:gfemunu}) is
given by:
\be\label{eqn:ourothree}
O_{\mu\nu ;\mu'\nu'}(x,x') = \frac{1}{2}(
g_{\mu\mu'} g_{\nu\nu'} + g_{\mu\nu'} g_{\nu\mu'}) \; ,
\ee
where we have defined bivectors:
\be\label{eqn:ourbivector}
g_{\mu\mu'}(x,x') = \partial_{\mu}\partial_{\mu'}u
-\frac{\partial_{\mu}u\partial_{\mu'}u}{2+u} \; .
\ee
The bitensor $O_{\mu\nu ;\mu'\nu'}(x,x')$ is determined by
requiring that it has the property:
\be
O_{\mu\nu ;\mu'\nu'}(x,x'=x)\,T^{\mu'\nu'}(x)=T_{\mu\nu}(x)
\; ,
\ee
which can be checked easily using (\ref{eqn:ourothree}),
(\ref{eqn:ourbivector}).

Apart from the $j$-dependent coupling constant $M_4^{(j)}$,
the expressions (\ref{eqn:fullgreenftn}) and (\ref{eqn:ourgone})
are the same as those derived by Naqvi \cite{Naqvi},\cite{D'Hoker}
for a massive symmetric tensor in $AdS_4$.
Note this is very different from the Green's function for the
DGP model, which in addition to an overall coupling constant
has extra gauge-dependent tensor structures \cite{Deffayet:2001uk}
which diverge as $M_P \to \infty$. In our straight gauge analysis
of warped DGP-like limits,
such effects are completely absent. The breakdown of
the linearized gravity approximation is due entirely to
strong coupling of the radion, as we discussed in the previous
section.

The coupling constant $M_4^{(j)}$ is given by
\be\label{eqn:gccfromgf}
\frac{1}{M_4^{(j)}{}^2} = \frac{H^2}{2kM^3} 
 \frac{b_L P_{p_j-1/2}^{-2}(-T_0) - a_L Q_{p_j-1/2}^2(-T_0)}
{[\,\partial_q(a_0 b_L - b_0 a_L)\,]_{q=q_j=p_j^2-\frac{9}{4}}}  \,.
\ee
This is the effective 4d gravitational coupling constant 
of the $j$-th KK mode of graviton to matter on the 0-brane.
Depending on the choice of input parameters, 
this coupling may show crossover behaviour; 
if the values of (\ref{eqn:gccfromgf}) for modes heavier than
some mass $\Lambda_{\rm co}$
are highly suppressed compared to lower lying modes, 
then $\Lambda_{\rm co}$ defines a crossover scale.

\subsection{asymptotic behavior of the Green's function}
The $u\to\infty$ limit of the graviton
Green's function (\ref{eqn:fullgreenftn}) shows how
sources on the 0-brane interact for geodesic separations
which are large compared to $1/H$, the $AdS_4$ radius of curvature.
The asymptotic formulae can be extracted from \cite{Bateman},
giving:
\be
G_{\mu\nu\,;\,\mu'\nu'}(x,x',y=y'=0) \gto{u\to\infty}
\sum_j G_1(u;p_j)
\bigl\{
\,T^{(3)}_{\mu\nu\,;\,\mu'\nu'} -
\frac{2H^4}{(p_j+\frac{1}{2})^2}u^2
T^{(1)}_{\mu\nu\,;\,\mu'\nu'} 
\bigr\}
\, ,
\ee
where
\be\label{eqn:ourasymp}
G_1(u;p_j) \gto{u\to\infty}
\frac{\sqrt{\pi}\,2^{-(p_j+\frac{1}{2})}}{4\pi^2 H^2 M_4^{(j)}{}^2}
\,\frac{\Gamma (p_j+3/2)}{\Gamma (p_j+1)}\,
u^{-(p_j+\frac{3}{2})}
\; .
\ee
For large $u$, $u \to {\rm exp}(H\mu )$. Thus from
(\ref{eqn:ourasymp}) we see that the contribution of
each graviton mode to the Euclidean Green's function damps exponentially
for large geodesic distances. In models with a $q < -9/4$
tachyon, $p_j$ is imaginary, and the asymptotic contribution to the
Green's function oscillates, as for a flat space tachyon.

\subsection{vDVZ discontinuity}
To investigate the vDVZ discontinuity problem, 
we will follow \cite{Porrati:2000cp}--\cite{Mouslopoulos:2002bk}.
In the flat 4d spacetime limit ($H\mu \ll 1$, \ie, $u\to 0$), 
expanding the Euclidean Green's function (\ref{eqn:fullgreenftn}) 
gives
\be\label{eqn:ourvdvz}
&&G_{\mu\nu\,;\,\mu'\nu'}(x,x',y=y'=0) \nn\\ 
&&\propto \sum_j\frac{1}{\mu^2} \Big( \delta_{\mu\mu'} \delta_{\nu\nu'} 
+ \delta_{\mu\nu'} \delta_{\nu'\nu'} 
- \frac{2}{3}\cdot\frac{3+q_j}{2+q_j} 
\delta_{\mu\nu} \delta_{\mu'\nu'} \Big) \,.
\ee
As $q_j$ varies from 0 to $\infty$, (\ref{eqn:ourvdvz}) varies
smoothly between the flat space Euclidean tensor structure of a massless
4d graviton:
\be
G_{\mu\nu\,;\,\mu'\nu'}^{\rm massless} \propto 
\frac{1}{\mu^2} (\delta_{\mu\mu'} \delta_{\nu\nu'} 
+ \delta_{\mu\nu'} \delta_{\nu\mu'} 
- \delta_{\mu\nu} \delta_{\mu'\nu'} ) \,,
\ee
and that of a massive 4d graviton:
\be
G_{\mu\nu\,;\,\mu'\nu'}^{\rm massless} \propto 
\frac{1}{\mu^2} (\delta_{\mu\mu'} \delta_{\nu\nu'} 
+ \delta_{\mu\nu'} \delta_{\nu\mu'} 
- \frac{2}{3}\delta_{\mu\nu} \delta_{\mu'\nu'} ) \,,
\ee
The condition for an ultralight graviton mode to avoid
the vDVZ discontinuity problem is that $q_j\to 0$
in the limit $H\to 0$.

\section{Warped models with an infinite extra dimension}

As an application of the different results developed in the previous
sections, we will highlight here 
the phenomenology of a
model with an infinite extra dimension. This generalizes the single
brane Karch-Randall (KR) model to include
a localized curvature term. 
In order to get $L=\infty$ we fix $w_L=12$ (thus $T_L=1$). 
Also we choose for
simplicity $v_L=0$. 
The model is therefore completely determined by
the higher dimensional Planck mass, $M$, the bulk curvature $k$, 
and the two brane parameters $v_0$ and $w_0$.
The ratio of brane to bulk curvature reads
\begin{equation}
\frac{H}{k}=\sqrt{1-T_0^2},
\end{equation}
where $T_0$ is given in terms of $v_0$ and $w_0$ 
by equation~(\ref{eqn:solfort}). 
Different limits of this model have been studied by a number of
authors~\cite{Pthesis},\cite{Mouslopoulos:2002bk}--\cite{Bazeia:2004yw}.

The infinite extra dimension results in the graviton
zero mode being non-normalizable and therefore it decouples from the
spectrum, whereas for strong warping (\textit{i.e.} large $k/H$) there
appears an ultralight ($m_1 \ll H$) graviton
mode. The ultralight mode couples to brane matter with comparable strength
as the RS zero mode, while the rest of the graviton
spectrum is heavier than $H$ and has much weaker couplings due
to the warping. For large localized curvature term, the mass of the
different modes decrease as we increase $v_0$, with the first mode
again becoming ultralight. The coupling of the KK modes to
brane matter, on the other 
hand, is suppressed only for modes heavier than some crossover scale.
This crossover scale depends on the size of $v_0$. In the following we shall 
see how our general results, when
particularized to this model, reproduce these features.

All our general
equations can be readily adapted to an infinite dimension by noticing that
$T_L=1$ implies $a_L=0$. 
The $y$-dependent part of the graviton KK mode wave function reads now
\begin{equation}
\mathcal{Y}^{(q)}(z)=P^{-2}_{(-1+\sqrt{9+4q})/2}(z),
\end{equation}
with $z\equiv \tanh k(y-y_0)$ and 
$k y_0=\tanh^{-1}T_0$. The masses are given by $m^2=q
H^2$, where $q$ is determined by
the zeroes of the equation
\begin{equation}
a_0(q)=0,
\end{equation}
with $a_0$ given in (\ref{eqn:ourasandbs}).
The kinetic coefficients for graviton KK modes and the radion
can be obtained directly from
(\ref{eqn:cgq}) and (\ref{eqn:radioncoeffi})
by taking the corresponding values of the different
parameters, 
\bea
\label{eqn:cgqKR}
{\mathcal C}^{(q)}_g &=& \frac{k}{H^2} \Big( 2\int_{-T_0}^{1}
{\mathcal Y}^{(q)}{}^2 dz  
+ v_0 (1-T_0^2) {\mathcal Y}^{(q)}{}^2(-T_0) \Big)\,, \\
\label{eqn:genradioncoeffiKR}
{\mathcal C}_\psi &=& \frac{3 \chi^2 H^2}{2 k} \frac{(1 + T_0)(2 
+ v_0 + v_0 T_0)}{1 + v_0 T_0} \,.
\eea
We have not written $C_g^{(0)}$ because, as expected, $T_L=1$ renders
the massless graviton
non-normalizable and 
therefore it decouples from the rest of the spectrum.
Similarly, the
couplings of the KK modes to the brane at $y=0$ can be obtained from
(\ref{eqn:gcq0}). 
\begin{figure}[ht]
  \begin{center}
    \resizebox{9cm}{!}{\includegraphics{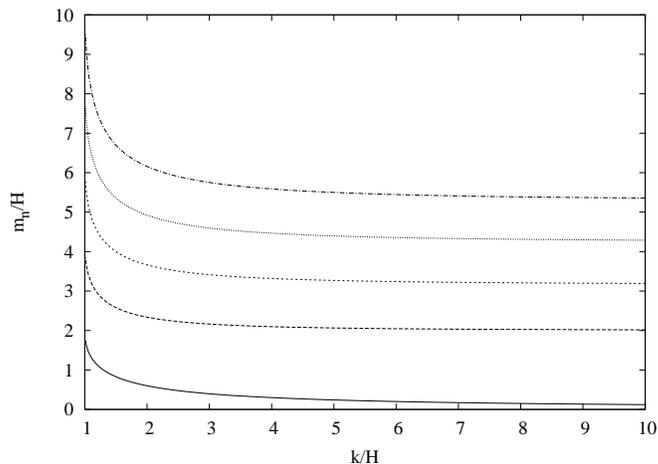}}
  \end{center}
  \caption{
Masses (in units of the brane curvature $H$) for the first five
massive graviton KK modes as a function of the bulk to brane curvature
ratio, $k/H$. The strong warping regime corresponds to the far right
region of the figure. 
  \label{mn:vs:Hbyk:plot}
}
\end{figure}
\begin{figure}[ht]
\centerline{
\hbox{
\resizebox{7cm}{!}{\includegraphics{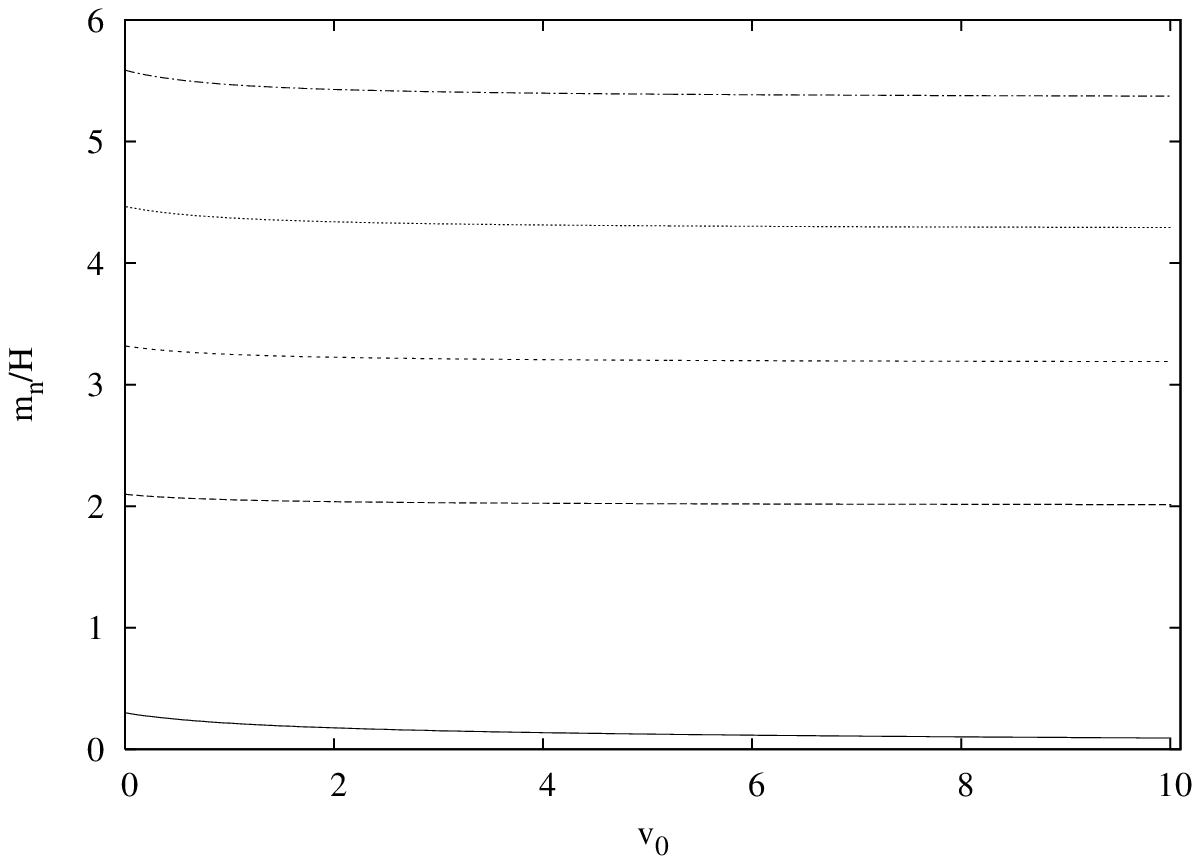}} 
\resizebox{7cm}{!}{\includegraphics{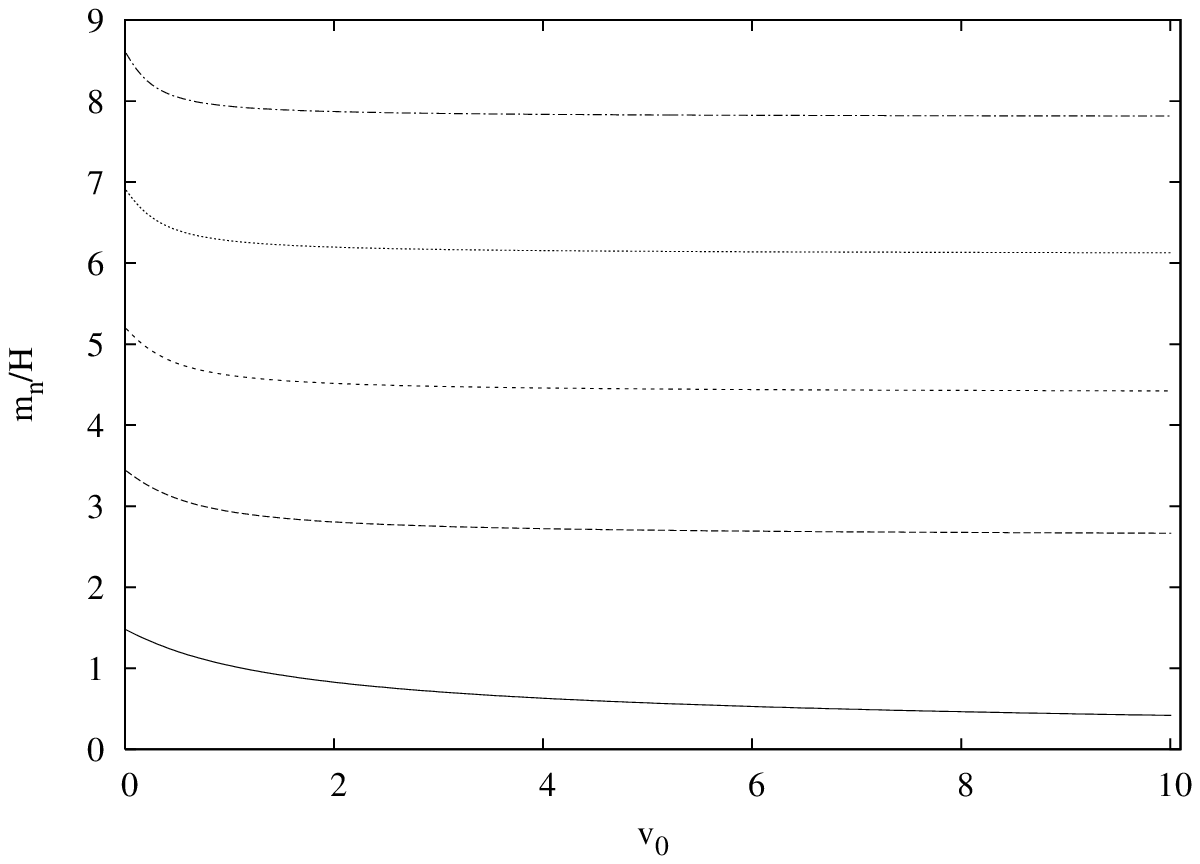}} 
}
}
  \caption{
Masses (in units of the brane curvature $H$) for the first five
massive graviton KK modes as a function of $v_0$ for fixed $H/k=0.25$
(left panel) and $H/k=0.95$ (right panel).
\label{mn:vs:v0:plot}
}
\end{figure}
As an example, we show in Figure~\ref{mn:vs:Hbyk:plot} 
the masses of the first five
modes (in units of the brane curvature $H$) as a function of $k/H$, for
$v_0=0$ (this is the original KR model). The warping
makes the mass of the different modes smaller as $k/H$ increases, and
in particular singles out the first massive mode as an ultralight
one, with mass much smaller than the brane curvature. In
Figure~\ref{mn:vs:v0:plot} we show the effect of the localized curvature
term by plotting the masses as a function of $v_0$
for fixed value of $H/k=0.25$ (left panel) and $H/k=0.95$ (right
panel). The effect is clearly not as dramatic as with the warping as
we can see by comparing the slopes of the curves with the ones of
Figure~\ref{mn:vs:Hbyk:plot} or the one on the right panel (small
warping) with the one on the left (large warping).

\begin{figure}[ht]
  \begin{center}
    \resizebox{9cm}{!}{\includegraphics{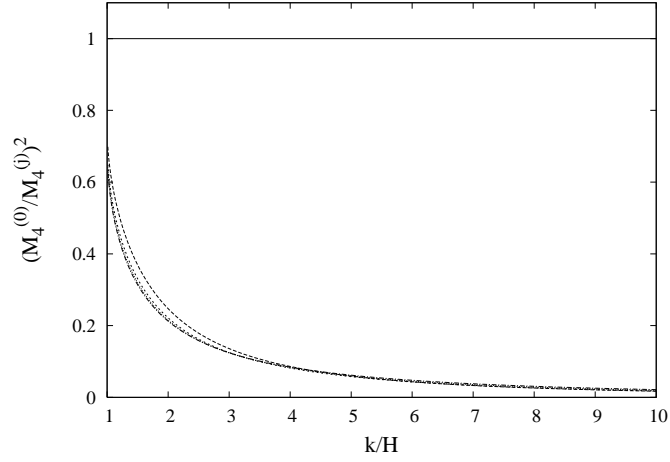}}
  \end{center}
  \caption{
Coupling of the first five massive graviton KK modes to the brane at
$y=0$ (in units of the coupling of the first mode)
as a function of the brane to bulk curvature
ratio, $H/k$. The strong warping regime corresponds to the left of the
figure. 
  \label{gn:vs:Hbyk:plot}
}
\end{figure}

\begin{figure}[ht]
\centerline{
\hbox{
\resizebox{7cm}{!}{\includegraphics{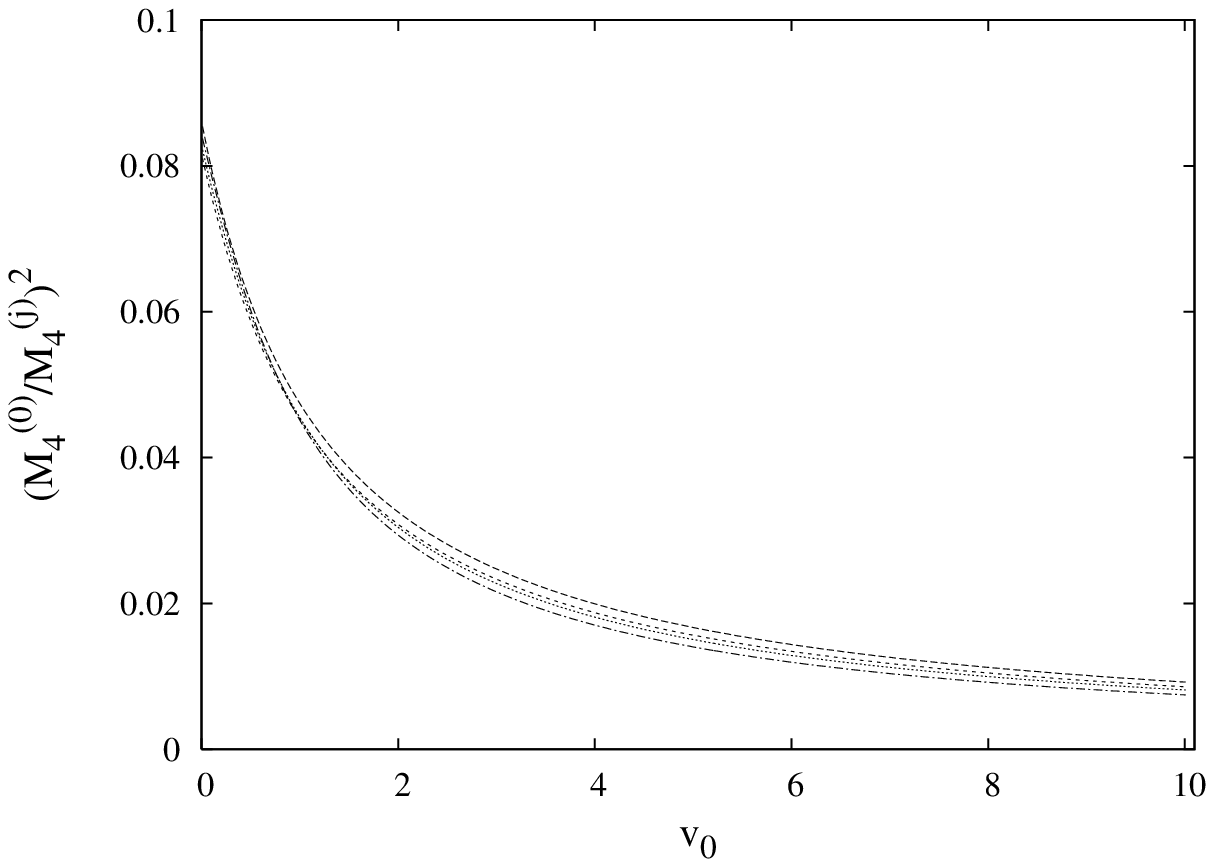}} 
\resizebox{7cm}{!}{\includegraphics{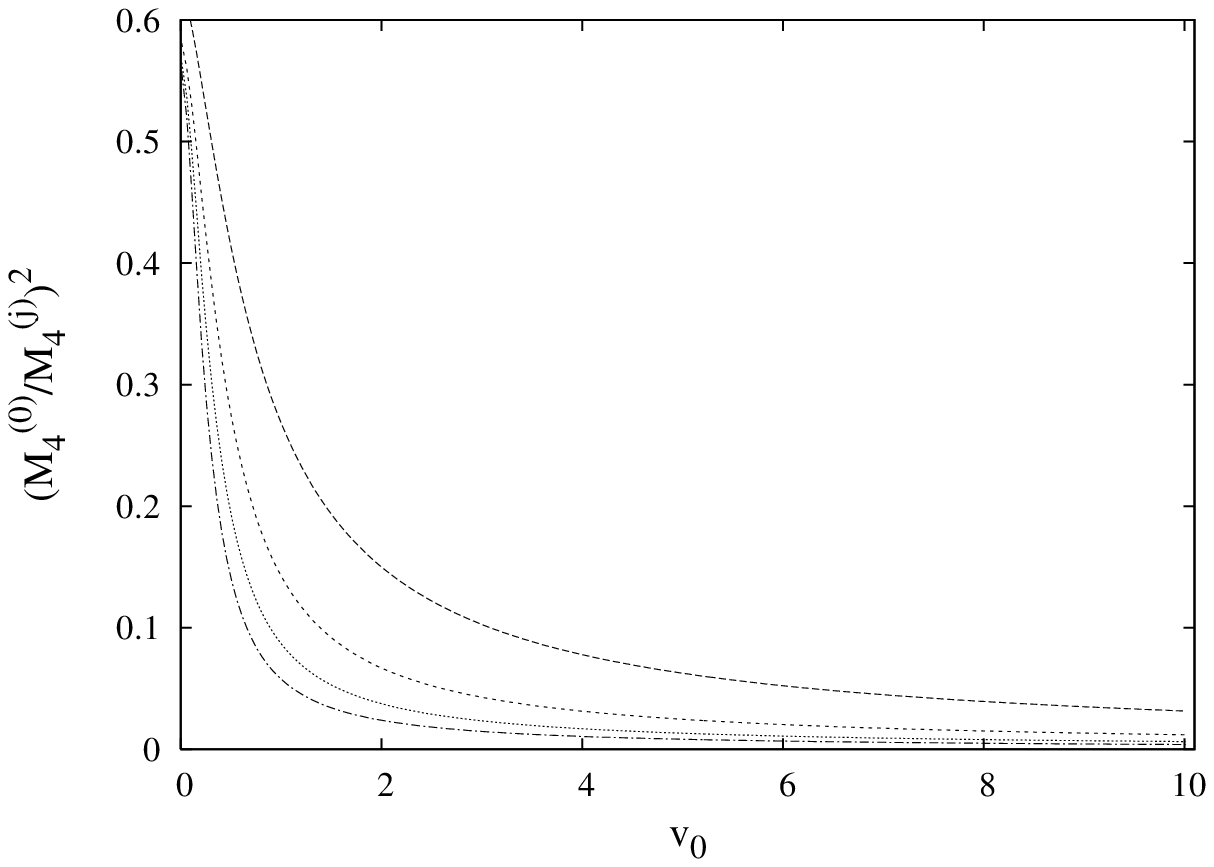}} 
}
}
\caption{
Coupling of the $j = 2$,3,4,5
massive graviton KK modes to the brane at
$y=0$ (in units of the coupling of the first mode)
as a function of $v_0$ for fixed $H/k=0.25$
(left panel) and $H/k=0.95$ (right panel).
\label{gn:vs:v0:plot}
}
\end{figure}

Now consider the couplings of the graviton modes to brane matter. In
Figure~\ref{gn:vs:Hbyk:plot} we show the couplings of the different KK modes to
the brane at $y=0$ in units of the first mode coupling. Again the warping
effect is clear, with very suppressed couplings for large warping
(large $k/H$). In Figure~\ref{gn:vs:v0:plot}, 
we show the coupling as a function of $v_0$, for fixed values of
$H/k=0.25$ (left) and $H/k=0.95$ (right). In the left panel we see how
the warping makes the effect of the curvature term less acute, whereas
the right one, where the warping is very small, clearly shows the
effect of $v_0$. This plot also shows that the
couplings of the different modes are more or less
suppressed depending on their masses (a clear hint of the crossover
behaviour). 

\begin{figure}[ht]
\centerline{\hbox{
\resizebox{9cm}{!}{\includegraphics{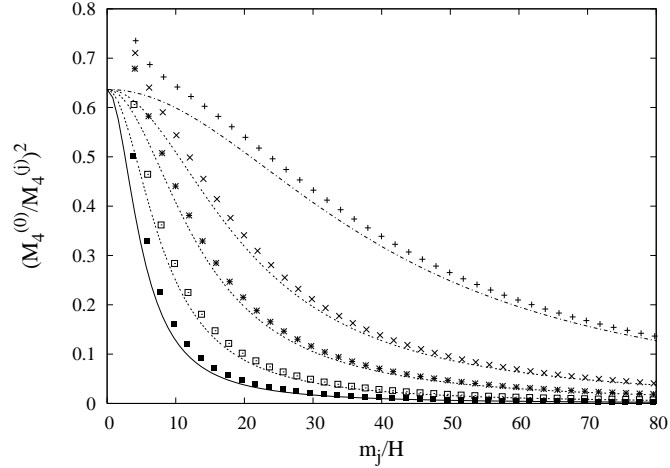}} 
}}
\caption{
The dots represent the 
couplings of the massive graviton KK modes to the brane at
$y=0$ (in units of the coupling of the first mode)
as a function of their mass (in units of $H$), 
for $v_0=0.05,0.1,0.15,0.25$ and
$0.4$, respectively and fixed $H/k=1$. The lines correspond to the
expectation from the DGP model with crossover scale $r_c=v_0/k$.
\label{gn:vs:mn:plot}
}
\end{figure}

The crossover behaviour is best seen by plotting the
coupling of the different modes as a function of their masses, for
different values of $v_0$. We do this in Figure~\ref{gn:vs:mn:plot}, 
where we have
fixed $H/k=1$ (small warping) and the
different sets of points correspond to $v_0=0.05, 0.1, 0.15, 0.25$ and
$0.4$, from top to
bottom. We have superimposed (solid lines) the expectation for these
couplings in the DGP model as given in~\cite{Dvali:2001gm}, 
with a crossover scale
$r_c=v_0/k$. It is clear from the
figure that once the localized curvature term is large enough to
suppress warping effects the agreement is quite good.

\begin{figure}[ht]
  \begin{center}
    \resizebox{9cm}{!}{\includegraphics{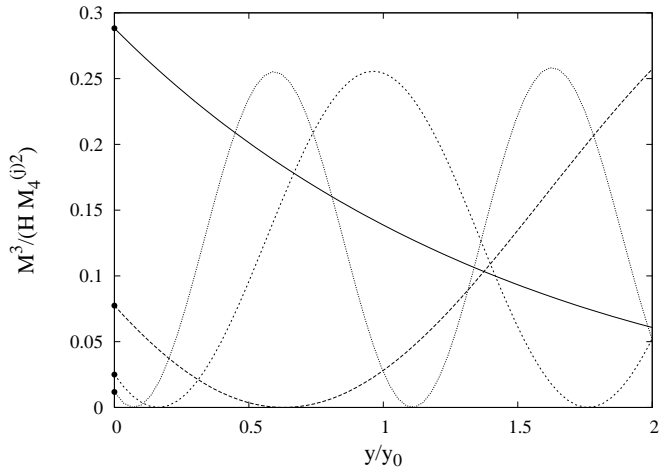}}
  \end{center}
  \caption{
Coupling of the $j-$th graviton KK mode (in units of $H/M^3$) 
to a probe brane localized at
$y$ for $0\leq y \leq 2 y_0$. We have fixed $H/k=0.95$ and $v_0=1$ and 
$j=1,2,4,6$ from top to bottom in the
coupling at the brane at $y=0$. The dots correspond to the couplings
to the brane as computed in the Green's function analysis.
  \label{gn2:vs:y:plot}
}
\end{figure}

Let us now turn to the Green's function analysis in the KR model.
Using the fact that $a_L=0$ we can write (\ref{eqn:closertofinalG1sol}) as,
\bea\label{eqn:getKaloper2}
&&\hspace{-25pt} G_1(u,y=y'=0) = \nn\\
&&\hspace{-25pt} \frac{i}{16\pi^3HM^3} \int_{-\infty}^\infty dp\,  
\frac{p\, P_{ip-1/2}^{-2}(-T_0)}{
-\frac{H}{2k} v_0 (p^2+\frac{9}{4}) P_{ip-1/2}^{-2}(-T_0) 
- P_{ip-1/2}^{-1}(-T_0)}  
\cdot \frac{Q_{ip-1/2}^1(1+u)}{\sqrt{u(2+u)}}  \,. 
\eea
Then we immediately see that our spectrum is determined by solving
\be
\frac{H}{2k} v_0 (p^2+\frac{9}{4}) P_{ip-1/2}^{-2}(-T_0) 
+ P_{ip-1/2}^{-1}(-T_0) = 0\,,
\label{poles:KR}
\ee
which of course gives the same result as the KK analysis.
We now contour-integrate (\ref{eqn:getKaloper2}) to get
\bea
\label{eqn:getKaloper3}
&&\hspace{-30pt} G_1(u,y=y'=0)=  \nn\\
&&\hspace{-30pt} \frac{1}{8\pi^2HM^3} \sum_{\nu > -1/2}
\Big[\frac{(\nu+\frac{1}{2})P_\nu^{-2}(-T_0)}{\partial_\nu 
\{ \frac{H}{2k} v_0 (\nu^2+\nu +\frac{5}{2}) P_\nu^{-2}(-T_0) 
+ P_\nu^{-1}(-T_0) \} }  
\cdot \frac{Q_\nu^1(1+u)}{\sqrt{u(2+u)}} \Big]_{\nu=p_j-\half} \,,
\eea
where the sum is over the poles of (\ref{poles:KR}).
Comparing this with (\ref{eqn:ourgone}), we obtain
\be\label{eqn:ourKRcoupling}
\frac{1}{M_4^{(j)}{}^2} = -\frac{H}{2M^3}
\Big[\frac{(\nu+\frac{1}{2})P_\nu^{-2}(-T_0)}{\partial_\nu 
\{ \frac{H}{2k} v_0 (\nu^2+\nu +\frac{5}{2}) P_\nu^{-2}(-T_0) 
+ P_\nu^{-1}(-T_0) \} } \Big]_{\nu=p_j-\half} \,.
\ee
As we show in Figure~\ref{gn2:vs:y:plot}, the couplings computed this
way (dots at $y=0$ in the figure) agree with the ones
computed in the KK analysis (lines
for couplings of the $j-$th KK modes, with $j=1,2,4,6$ to a probe
brane as a function of its location). We have fixed $H/k=0.95$ and
$v_0=1$ in this plot and consider values of $y\leq 2 y_0$. It is
also trivial to check that the masses and couplings agree with the
results in~\cite{Kaloper:2005wq}, by taking $v_0=0$ in the equations
above.   

\section{Conclusion}

One of the frustrations of trying to understand
braneworld gravity is the large number of models which display
various combinations of interesting or problematic features.
In this paper, following \cite{clp}, we have addressed this
difficulty by introducing a general unified framework.
This framework describes general 5d warped setups with two
branes and localized curvature. Apart from the
5d Planck mass $M$ and the bulk curvature $k$, there are only four input
parameters: $(v_0, w_0, v_L, w_L)$, related to the two brane
tensions and the strength of the localized curvature.
This framework includes as special
cases both Randall-Sundrum models, both Karch-Randall models,
the Lykken-Randall model, bigravity models, as well as 
a large class of DGP-like models, where the localized curvature
competes with or dominates the warping.

Our unified framework has the additional advantage that we
can gauge-fix the 5d general covariance explicitly, and exhibit
the gauge-fixed 4d effective action for the physical
degrees of freedom. This we have done in section 2, building
on the results of \cite{clp}. In section 3, we used this effective
action, together with an analysis of the KK graviton spectrum,
to reveal the presence of kinetic ghosts and tachyons in
a large class of models. Depending on choices of our four
input parameters, we found examples of models where the
radion is a ghost, the massless graviton is a ghost, or
the massive graviton modes are ghosts. We also found models containing
one or two tachyons in the graviton spectrum.

Our framework is especially well-suited for examining the
problem of strong coupling. In the literature, the
phenomenon we call strong coupling has shown up in many
confusing guises, all having to do with a breakdown of
the linearized approximation for braneworld gravity. Here,
we mapped out the parameter space in which either the radion
or the massless graviton become strongly coupled. We found
that the graviton only becomes strongly coupled for models
which also contain tachyons and ghosts. We also noted that,
in our straight gauge formalism, the graviton propagator
does not have tensor structures which diverge as an
ultralight mode becomes massless \cite{Deffayet:2001uk}.
Because we are always in a
straight gauge, we also don't have to worry about large
brane-bending effects \cite{Porrati:2002cp}.

Thus in our framework strong coupling is an isolated property
of the radion, and can be studied in a straightforward way.
In section 4 we have written a general formula for the
resulting strong coupling scale in our framework. We observed
that there is a tight relationship between strong coupling
and the presence of an ultralight graviton mode. This was
known for the DGP and bigravity models, but here we see it
in a unified picture that interpolates smoothly between
both kinds of models. As expected, the vDVZ discontinuity
is never present for ultralight modes in our warped setups.

In section 6 we studied the effects of localized curvature
in the subset of models that have an infinite extra dimension.
These are the models in which the usual 4d massless graviton
is absent, having decoupled in the limit that we removed the
second brane.
We saw that localized curvature and warping both have
the effect of making an ultralight graviton from the first
massive KK mode. We compared these two effects in models
which contain both. The dimensionless parameter $v_0$
controls the strength of the localized curvature, while
the dimensionless ratio $k/H$ measures the strength of
the warping. By looking at models with comparable values
for these parameters, we saw that the warping effect is stronger.
Thus ``locally localized'' gravity is more efficient than
localized curvature in creating a simulacrum of 4d gravity.

Finally, in section 6 we exhibited explicit crossover
behavior for models in our general setup. In these models
the couplings of the KK graviton modes to brane matter are
unsuppressed up to a mass scale $1/r_c \sim k/v_0$, and are
highly suppressed for modes heavier than this scale.
This behavior is the most interesting phenomenological
feature of DGP braneworld gravity; here we have found it
in a general warped framework. Our results, as seen in
Figure~\ref{gn:vs:mn:plot}, agree quantitatively with
what one would predict by analogy with DGP.

We have opened a window to a new class of models
which have crossover scales from four
dimensional gravity at short distance to five
dimensional gravity at longer distances. We have presented
a well-defined framework to analyze strong coupling behavior,
which remains a fundamental obstacle towards developing
realistic effective theories of braneworld gravity.

\subsection*{Acknowledgments}
We thank Eduardo Pont\'on, Michele Redi,
and Robert Wald for useful discussions.
This research was supported by the U.S.~Department of Energy
Grants DE-AC02-76CHO3000 and DE-FG02-90ER40560.

\newpage

\appendix
\section{Radion action to cubic order}\label{appA}
With $h_{\mu4} = 0$, the general cubic expansion of (\ref{eqn:ouraction}) is
\bea\label{eqn:gen3action}
&&\frac{S}{2M^3}\Big|_{\rm cubic} \nn\\
&&= \int d^5x \sqrt{-g} \Big( \frac{H^2}{4a^2} \tilde h^3 
- \frac{3H^2 + a'{}^2}{2a^2} \tilde h h^{\mu\nu} h_{\mu\nu} 
+ 2k^2 h^{\mu\nu} h_{\mu\alpha} h_\nu^\alpha \nn\\
&&\quad + \frac{1}{8} \tilde h \tilde h'{}^2 
+ \frac{1}{2} h^{\mu\nu}{}' h_{\mu\alpha}' h_\nu^\alpha 
- \frac{1}{8} \tilde h h^{\mu\nu}{}' h_{\mu\nu}' 
- \frac{1}{4} \tilde h' (h^{\mu\nu} h_{\mu\nu})' \nn\\
&&\quad - \frac{1}{16} \tilde h^2 \tn^2 \tilde h 
+ \frac{1}{4} \tilde h h^{\mu\nu} \tn_\mu \tn_\nu \tilde h 
- \frac{1}{8} \tilde h \tn_\mu h^{\alpha\beta} \tn^\mu h_{\alpha\beta} 
- \frac{1}{4} \tn_\mu \tilde h \tn^\mu (h^{\alpha\beta} h_{\alpha\beta}) \nn\\
&&\quad - \frac{1}{2} \tilde h \tn_\mu (h^{\mu\alpha} \tn_\nu h^\nu_\alpha) 
- \frac{1}{4} \tilde h \tn_\nu h^{\mu\alpha} \tn_\mu h^\nu_\alpha
- \frac{1}{2} \tilde h h^{\mu\alpha} \tn_\nu \tn_\mu h^\nu_\alpha \nn\\
&&\quad + \frac{1}{4} h^{\mu\nu} \tn_\mu h^{\alpha\beta} 
\tn_\nu h_{\alpha\beta} 
- \frac{1}{4} h^{\mu\nu} \tn_\mu \tn_\nu (h^{\alpha\beta} h_{\alpha\beta})
+ \frac{1}{2} h^{\mu\nu} \tn_\alpha h_{\mu\beta} \tn^\alpha h^\beta_\nu \nn\\
&&\quad - h^{\mu\nu} \tn_\mu h^{\alpha\beta} \tn_\beta h_{\nu\alpha}
- \frac{1}{2} h^{\mu\nu} \tn_\alpha h^\beta_\mu \tn_\beta h^\alpha_\nu \nn\\
&&\quad - \frac{3a'{}^2}{a^2} h_{44}^3
+ \frac{-3H^2 + 12a'{}^2}{8a^2} \tilde h h_{44}^2 
+ \frac{9a'}{8a} \tilde h' h_{44}^2 \nn\\
&&\quad + \frac{3H^2}{4a^2} \tilde h^2 h_{44} 
+ \frac{-3H^2 + a'{}^2}{2a^2} h^{\alpha\beta} h_{\alpha\beta} h_{44} \nn\\
&&\quad - \frac{3a'}{8a} \tilde h^2{}' h_{44} 
- \frac{1}{8} \tilde h'{}^2 h_{44} 
+ \frac{1}{8} h^{\alpha\beta}{}' h_{\alpha\beta}' h_{44}
+ \frac{3a'}{4a} (h^{\alpha\beta} h_{\alpha\beta})' h_{44} \nn\\
&&\quad + \frac{1}{8} h_{44}^2 \tn^2 \tilde h 
- \frac{1}{8} h_{44}^2 \tn_\mu \tn_\nu h^{\mu\nu} 
- \frac{1}{16} \tilde h^2 \tn^2 h_{44} 
- \frac{1}{8} h_{44} \tilde h \tn^2 \tilde h 
+ \frac{1}{4} \tilde h h^{\mu\nu} \tn_\mu \tn_\nu h_{44} \nn\\
&&\quad + \frac{1}{4} h_{44} h^{\mu\nu} \tn_\mu \tn_\nu \tilde h 
- \frac{1}{8} h_{44} \tn_\mu h^{\alpha\beta} \tn^\mu h_{\alpha\beta} 
- \frac{1}{4} \tn_\mu h_{44} \tn^\mu (h^{\alpha\beta} h_{\alpha\beta}) \nn\\
&&\quad - \frac{1}{2} h_{44} \tn_\mu (h^{\mu\alpha} \tn_\nu h^\nu_\alpha) 
- \frac{1}{4} h_{44} \tn_\nu h^{\mu\alpha} \tn_\mu h^\nu_\alpha
- \frac{1}{2} h_{44} h^{\mu\alpha} \tn_\nu \tn_\mu h^\nu_\alpha \Big) \nn\\
&&+ \sum_i \int d^4 x \Big[ \sqrt{-g} \, 
\lambda_i \Big( \frac{H^2}{4a^2} \tilde h^3 
- \frac{3H^2}{2a^2} \tilde h h^{\mu\nu} h_{\mu\nu} 
+ \frac{2H^2}{a^2} h^{\mu\nu} h_{\mu\alpha} h_\nu^\alpha \nn\\
&&\quad - \frac{1}{16} \tilde h^2 \tn^2 \tilde h 
+ \frac{1}{4} \tilde h h^{\mu\nu} \tn_\mu \tn_\nu \tilde h 
- \frac{1}{8} \tilde h \tn_\mu h^{\alpha\beta} \tn^\mu h_{\alpha\beta} 
- \frac{1}{4} \tn_\mu \tilde h \tn^\mu (h^{\alpha\beta} h_{\alpha\beta}) \nn\\
&&\quad - \frac{1}{2} \tilde h \tn_\mu (h^{\mu\alpha} \tn_\nu h^\nu_\alpha) 
- \frac{1}{4} \tilde h \tn_\nu h^{\mu\alpha} \tn_\mu h^\nu_\alpha
- \frac{1}{2} \tilde h h^{\mu\alpha} \tn_\nu \tn_\mu h^\nu_\alpha \nn\\
&&\quad + \frac{1}{4} h^{\mu\nu} 
\tn_\mu h^{\alpha\beta} \tn_\nu h_{\alpha\beta} 
- \frac{1}{4} h^{\mu\nu} \tn_\mu \tn_\nu (h^{\alpha\beta} h_{\alpha\beta})
+ \frac{1}{2} h^{\mu\nu} \tn_\alpha h_{\mu\beta} \tn^\alpha h^\beta_\nu \nn\\
&&\quad - h^{\mu\nu} \tn_\mu h^{\alpha\beta} \tn_\beta h_{\nu\alpha}
- \frac{1}{2} h^{\mu\nu} \tn_\alpha h^\beta_\mu 
\tn_\beta h^\alpha_\nu \Big) \Big]_{y=y_i}\,.
\eea
The following identities will be useful:
\bea
&&\nabla_\mu h^{\alpha\beta} = \tn_\mu h^{\alpha\beta}\,,
\quad \nabla_\mu h_{44} = \tn_\mu h_{44}\,,\nn\\
&&\nabla_\mu h^{\alpha4} = \frac{a'}{a} (\delta^\alpha_\mu h_{44} 
- h^\alpha_\mu)\,, \quad
\nabla_4 h^{\alpha4} = 0\,, \nn\\
&&\nabla_4 h^{\alpha\beta} = h^{\alpha\beta}{}' 
+ \frac{2a'}{a} h^{\alpha\beta} \,,\quad
\nabla_4 h_{\alpha\beta} = h_{\alpha\beta}' 
- \frac{2a'}{a} h^{\alpha\beta} \,, \nn\\
&&h^{\alpha\beta}{}' h_{\alpha\beta} 
= \frac{1}{2}(h^{\alpha\beta} h_{\alpha\beta})' 
- \frac{2a'}{a} h^{\alpha\beta} h_{\alpha\beta} \,, \quad
h^{\alpha\beta} h_{\alpha\beta}' 
= \frac{1}{2}(h^{\alpha\beta} h_{\alpha\beta})' 
+ \frac{2a'}{a} h^{\alpha\beta} h_{\alpha\beta} \,, \nn\\
&&h^{\alpha\beta}{}' h_{\alpha\mu} h^\mu_\beta
= \frac{1}{3}(h^{\alpha\beta} h_{\alpha\mu} h^\mu_\beta)' 
- \frac{2a'}{a} h^{\alpha\beta} h_{\alpha\mu} h^\mu_\beta \,, \nn\\ 
&&h^{\alpha\beta} h_{\alpha\mu}' h^\mu_\beta
= \frac{1}{3}(h^{\alpha\beta} h_{\alpha\mu} h^\mu_\beta)' 
+ \frac{2a'}{a} h^{\alpha\beta} h_{\alpha\mu} h^\mu_\beta \,. \nn
\eea
Upon making the substitutions
\be
h_{\mu\nu} \to a^2 {\mathcal Y}_1(y) \tn_\mu \tn_\nu \psi 
+ g_{\mu\nu} {\mathcal Y}_2(y) \psi 
\; , \quad h_{44} = F \psi\,, 
\ee
(\ref{eqn:gen3action}) produces a few hundred terms. 
Collecting terms with three ${\mathcal Y}_1$'s, the bulk part gives
\bea\label{eqn:r3a6}
&&a^6 \Big\{ {\mathcal Y}_1^3 \Big( -\frac{1}{16} (\tn^2 \psi)^2 \tn^4 \psi 
+ \frac{1}{4} \tn^2 \psi \tn^\mu \tn^\nu \psi \tn_\mu \tn_\nu \tn^2 \psi 
- \frac{1}{8} \tn^2 \psi \tn_\mu \tn_\alpha 
\tn_\beta \psi \tn^\mu \tn^\alpha \tn^\beta \psi \nn\\
&&\qquad - \frac{1}{4} \tn_\mu \tn^2 \psi 
\tn^\mu (\tn_\alpha \tn_\beta \psi \tn_\alpha \tn_\beta \psi) 
- \frac{1}{2} \tn^2 \psi \tn_\mu (\tn^\mu 
\tn^\alpha \psi \tn^2 \tn_\alpha \psi) \nn\\
&&\qquad - \frac{1}{4} \tn^2 \psi \tn_\mu 
\tn_\alpha \tn_\beta \psi \tn^\alpha \tn^\mu \tn^\beta \psi 
- \frac{1}{2} \tn^2 \psi \tn^\mu \tn^\alpha 
\psi \tn_\nu \tn_\mu \tn^\nu \tn_\alpha \psi \nn\\
&&\qquad + \frac{1}{4} \tn^\mu \tn^\nu \psi 
\tn_\mu \tn_\alpha \tn_\beta \psi \tn_\nu \tn^\alpha \tn^\beta \psi 
- \frac{1}{4} \tn^\mu \tn^\nu \psi \tn_\mu 
\tn_\nu (\tn_\alpha \tn_\beta \psi \tn^\alpha \tn^\beta \psi) \nn\\
&&\qquad + \frac{1}{2} \tn^\mu \tn^\nu \psi 
\tn_\alpha \tn_\beta \tn_\mu \psi \tn^\alpha \tn^\beta \tn_\nu \psi 
- \tn^\mu \tn^\nu \psi \tn_\mu \tn_\alpha 
\tn_\beta \psi \tn^\alpha \tn^\beta \tn_\nu \psi \\
&&\qquad - \frac{1}{2} \tn^\mu \tn^\nu \psi 
\tn_\alpha \tn_\beta \tn_\mu \psi \tn^\beta \tn^\alpha \tn_\nu \psi \nn\\
&&\qquad + \frac{H^2}{4a^2} (\tn^2 \psi)^3 
- \frac{3H^2}{2a^2} \tn^2 \psi \tn_\alpha 
\tn_\beta \psi \tn^\alpha \tn^\beta \psi 
+ \frac{2H^2}{a^2} \tn_\mu \tn_\nu \psi 
\tn_\alpha \tn^\mu \psi \tn^\alpha \tn^\nu \psi \Big) \nn\\
&&\quad + {\mathcal Y}_1 {\mathcal Y}_1'{}^2 \Big( \frac{1}{8} (\tn^2 \psi)^3 
- \frac{5}{8} \tn^2 \psi \tn_\alpha \tn_\beta \psi \tn^\alpha \tn^\beta \psi 
+ \frac{1}{2} \tn_\mu \tn_\nu \psi 
\tn_\alpha \tn^\mu \psi \tn^\alpha \tn^\nu \psi \Big) \Big\}\,. \nn
\eea
Performing 4d integration by parts 
a few dozen times, a dramatic simplification occurs 
and (\ref{eqn:r3a6}) becomes:
\bea
&&a^6 {\mathcal Y}_1 {\mathcal Y}_1'{}^2 \Big\{ - \frac{1}{8} (\tn^2 \psi)^3 
+ \frac{1}{8} \tn^2 \psi \tn_\alpha 
\tn_\beta \psi \tn^\alpha \tn^\beta \psi \nn\\
&&\qquad + \frac{H^2}{a^2} \Big( \psi (\tn^2 \psi)^2 
- \psi \tn_\alpha \tn_\beta \psi \tn^\alpha \tn^\beta \psi 
- \frac{3H^2}{2a^2} \psi^2 \tn^2 \psi \Big) \Big\} \,.
\eea
By a similar calculation, we can show that no term with 
three ${\mathcal Y}_1$'s survives 
in the brane-boundary part. Note that terms with 
eight $\tn$'s are all cancelled.

Repeating the above procedure with the remaining terms, 
we get the radion cubic action:
\bea
\frac{S}{2M^2}\Big|_{\psi^3} 
&=& 2 \int d^4 x \int_0^L dy \sqrt{-g} 
\Big\{ a^6 {\mathcal Y}_1 {\mathcal Y}_1'{}^2 \Big\{ 
- \frac{1}{8} (\tn^2 \psi)^3 
+ \frac{1}{8} \tn^2 \psi \tn_\alpha \tn_\beta \psi 
\tn^\alpha \tn^\beta \psi \nn\\
&&\qquad + \frac{H^2}{a^2} \Big( \psi (\tn^2 \psi)^2 
- \psi \tn_\alpha \tn_\beta \psi \tn^\alpha \tn^\beta \psi 
- \frac{3H^2}{2a^2} \psi^2 \tn^2 \psi \Big) \Big\} \nn\\
&& + a^4 {\mathcal Y}_1^2 {\mathcal Y}_2 \Big\{ \frac{1}{8} (\tn^2 \psi)^3 
- \frac{1}{8} \tn^2 \psi \tn_\alpha \tn_\beta \psi 
\tn^\alpha \tn^\beta \psi \nn\\
&&\qquad + \frac{H^2}{a^2} \Big( - \frac{3}{4} 
\psi (\tn^2 \psi)^2 + \frac{9H^2}{4a^2} \psi^2 \tn^2 \psi \Big) \Big\} \nn\\
&& + a^4 {\mathcal Y}_1 {\mathcal Y}_1' {\mathcal Y}_2' 
\Big( \frac{1}{4} \psi (\tn^2 \psi)^2 
- \psi \tn_\alpha \tn_\beta \psi \tn^\alpha \tn^\beta \psi \Big) 
+ \Big( \frac{3H^2}{2} {\mathcal Y}_1 {\mathcal Y}_2^2 
+ \frac{3}{4} {\mathcal Y}_2^3 \Big) \psi^2 \tn^2 \psi \nn\\
&& + a^4 F {\mathcal Y}_1^2 \Big\{ \frac{1}{16} (\tn^2 \psi)^3 
- \frac{1}{16} \tn^2 \psi \tn_\alpha \tn_\beta 
\psi \tn^\alpha \tn^\beta \psi \nn\\
&&\qquad + \frac{H^2}{a^2} \Big( - \frac{3}{8} 
\psi (\tn^2 \psi)^2 + \frac{9H^2}{8a^2} \psi^2 \tn^2 \psi \Big) \Big\} \nn\\
&& + a^4 F {\mathcal Y}_1 {\mathcal Y}_1' \frac{3a'}{4a} 
(- \psi (\tn^2 \psi)^2 + 2 \psi \tn_\alpha 
\tn_\beta \psi \tn^\alpha \tn^\beta \psi ) \nn\\
&& + \frac{1}{8} a^4 F {\mathcal Y}_1'{}^2 
(- \psi (\tn^2 \psi)^2 + \psi \tn_\alpha 
\tn_\beta \psi \tn^\alpha \tn^\beta \psi ) \\
&& + a^2 F {\mathcal Y}_1 {\mathcal Y}_2 
\Big( -\frac{1}{2} \psi (\tn^2 \psi)^2 + 
\frac{1}{2} \psi \tn_\alpha \tn_\beta \psi \tn^\alpha \tn^\beta \psi 
+ \frac{3H^2}{a^2} \psi^2 \tn^2 \psi \Big) \nn\\
&& - a^2 F \Big( \frac{3a'}{2a} ({\mathcal Y}_1' 
{\mathcal Y}_2 + {\mathcal Y}_1 {\mathcal Y}_2') 
+ \frac{3}{4} {\mathcal Y}_1' {\mathcal Y}_2' \Big) \psi^2 \tn^2 \psi \nn\\
&& + F {\mathcal Y}_2^2 \Big( -\frac{3}{8} \psi^2 \tn^2 \psi 
+ \frac{6H^2}{a^2} \psi^3 \Big) 
- F \Big( \frac{6a'}{a} {\mathcal Y}_2 {\mathcal Y}_2' 
+ \frac{3}{2} {\mathcal Y}_2'{}^2 \Big) \psi^3 \nn\\
&& + a^2 F^2 \Big( \frac{3a'{}^2}{2a^2} {\mathcal Y}_1 
+ \frac{9a'}{8a} {\mathcal Y}_1' \Big) \psi^2 \tn^2 \psi \nn\\
&& + F^2 {\mathcal Y}_2 \Big( \frac{3}{8} \psi^2 \tn^2 \psi 
+ \frac{-3H^2 + 12a'{}^2}{2a^2} \psi^3 \Big) 
+ F^2 {\mathcal Y}_2' \frac{9a'}{2a} \psi^3 
- \frac{3a'{}^2}{a^2} F^3 \psi^3 \Big\} \nn\\
&& + \sum_i \int d^4 x \Big[ \sqrt{-g} 
\Big\{ a^4 {\mathcal Y}_1^2 {\mathcal Y}_2 \Big\{ \frac{1}{8} (\tn^2 \psi)^3 
- \frac{1}{8} \tn^2 \psi \tn_\alpha 
\tn_\beta \psi \tn^\alpha \tn^\beta \psi \nn\\
&&\qquad\qquad + \frac{H^2}{a^2} \Big( 
- \frac{3}{4} \psi (\tn^2 \psi)^2 + \frac{9H^2}{4a^2} 
\psi^2 \tn^2 \psi \Big) \Big\} \nn\\
&&\qquad + \Big( \frac{3H^2}{2} {\mathcal Y}_1 {\mathcal Y}_2^2 
+ \frac{3}{4} {\mathcal Y}_2^3 \Big) \psi^2 \tn^2 
\psi \nn \Big\} \Big]_{y=y_i} \,.
\eea  
Collecting terms of the same $\tn$-structure, we get (\ref{eqn:r3}).

\section{Graviton Green's function}\label{appB}
In this section, we explicitly calculate the graviton Green's 
function in the Euclidean signature. 

In \cite{clp}, the bulk EOM for the graviton and the radion are obtained to be
\bea\label{eqn:bulkmunueom2}
0 &=& \tilde\nabla_\rho \tilde\nabla_\mu h^\rho_\nu 
+ \tilde\nabla_\rho \tilde\nabla_\nu h^\rho_\mu 
- \tilde\nabla^2 h_{\mu\nu} - \tilde\nabla_\mu \tilde\nabla_\nu \tilde h 
- g_{\mu\nu} (\tilde\nabla_\rho \tilde\nabla_\sigma h^{\rho\sigma} 
- \tilde\nabla^2 \tilde h) \nn\\
&& - h_{\mu\nu}'' + g_{\mu\nu} \tilde h'' 
+ \frac{4a'}{a} g_{\mu\nu} \tilde h'
+ \frac{8H^2 + 4a'{}^2}{a^2} h_{\mu\nu} 
- \frac{3H^2}{a^2} g_{\mu\nu} \tilde h \nn\\
&& - F \tilde\nabla_\mu \tilde\nabla_\nu \psi +
g_{\mu\nu} F \tilde\nabla^2 \psi 
- \frac{3a'}{a} g_{\mu\nu} F' \psi - \frac{6H^2 
+ 12 a'{}^2}{a^2} g_{\mu\nu} F \psi \,,\\
\label{eqn:bulkmu4eom2}
0 &=& (\tilde\nabla_\nu h^\nu_\mu)' - \partial_\mu \tilde h' 
+ \frac{3a'}{a} F \partial_\mu \psi \,,\\
\label{eqn:bulk44eom2}
0 &=& -\tilde\nabla_\mu \tilde\nabla_\nu h^{\mu\nu} 
+ \tilde\nabla^2 \tilde h + \frac{3a'}{a} \tilde h' 
- \frac{3H^2}{a^2} \tilde h - \frac{12 a'{}^2}{a^2} F \psi \,, \\
0 &=& \frac{(a^2 \tilde h')'}{a^2} 
+ F \tilde\nabla^2 \psi - \frac{4a'}{a} F' \psi - 8k^2 F \psi\,,
\eea
and the brane-boundary equation is
\bea\label{eqn:bdyeom1}
0 &=& \Big[\theta_i (h_{\mu\nu}{}' - g_{\mu\nu} \tilde h') 
+ \Big(\frac{3\lambda_i H^2}{a^2} - 2k T_i \Big) h_{\mu\nu} 
- \frac{3\lambda_i H^2}{2a^2} g_{\mu\nu}\tilde h 
+ 3kT_i g_{\mu\nu} F \psi \nn\\
&&+ \frac{\lambda_i}{2}(\tilde\nabla_\rho \tilde\nabla_\mu h^\rho_\nu 
+ \tilde\nabla_\rho \tilde\nabla_\nu h^\rho_\mu 
- \tilde\nabla^2 h_{\mu\nu} 
- \tilde\nabla_\mu \tilde\nabla_\nu \tilde h) \nn\\
&&- \frac{\lambda_i}{2} g_{\mu\nu} 
(\tilde\nabla_\rho \tilde\nabla_\sigma h^{\rho\sigma} 
- \tilde\nabla^2 \tilde h) \Big]_{y=y_i}\,.
\eea 
Using (\ref{eqn:bulkmu4eom2}) or (\ref{eqn:bulk44eom2}) we 
eliminate the radion 
and get the equations involving the graviton only. The bulk 
part gives only two independent 
equations:
\bea\label{eqn:gravonlyeq1}
0 &=& \tilde\nabla_\mu \tilde\nabla_\rho h^\rho_\nu 
+ \tilde\nabla_\nu \tilde\nabla_\rho h^\rho_\mu 
- \tilde\nabla^2 h_{\mu\nu} - \tilde\nabla_\mu \tilde\nabla_\nu \tilde h 
+ g_{\mu\nu} \Big(-\frac{2}{3}\tilde\nabla_\rho 
\tilde\nabla_\sigma h^{\rho\sigma} 
+ \frac{1}{2} \tilde\nabla^2 \tilde h\Big) \nn\\
&& - h_{\mu\nu}'' + g_{\mu\nu} \frac{1}{4} \tilde h'' 
+ \frac{a'}{a} g_{\mu\nu} \tilde h'
+ \frac{4a'{}^2}{a^2} h_{\mu\nu} + \frac{H^2}{2a^2} g_{\mu\nu} \tilde h \\
&& + \frac{a}{3a'}\Big(\frac{1}{2}(\tn_\mu\tn_\rho h^\rho_\nu{}' 
+ \tn_\nu\tn_\rho h^\rho_\mu{}') - \tn_\mu\tn_\nu \tilde h' \Big) 
+ g_{\mu\nu} \frac{a}{12a'}(\tn^2 \tilde h' - \tn_\rho 
\tn_\sigma h^{\rho\sigma}{}') \,, \nn\\
\label{eqn:gravonlyeq2}
0 &=& -\tn_\mu\tn_\rho\tn_\sigma h^{\rho\sigma} 
+ \tn_\mu\tn^2 \tilde h - \frac{a'}{a}\tn_\mu \tilde h' 
- \frac{3H^2}{a^2}\tn_\mu \tilde h + \frac{4a'}{a}\tn_\nu h^\nu_\mu{}' \,,
\eea
and the brane-boundary part gives
\bea\label{eqn:gravonlybbeq}
0 &=& \Big[\theta_i (h_{\mu\nu}{}' - g_{\mu\nu} \frac{1}{4}\tilde h') 
- \Big(\frac{\lambda_i H^2}{a^2} + 2k T_i \Big) h_{\mu\nu} 
- \frac{H^2}{2a^2} \Big(\lambda_i + \frac{3}{2kT_i} \Big) 
g_{\mu\nu}\tilde h \nn\\
&&+ \frac{\lambda_i}{2}(\tilde\nabla_\mu \tilde\nabla_\rho h^\rho_\nu 
+ \tilde\nabla_\nu \tilde\nabla_\rho h^\rho_\mu 
- \tilde\nabla^2 h_{\mu\nu} 
- \tilde\nabla_\mu \tilde\nabla_\nu \tilde h) \nn\\
&&- \Big(\frac{\lambda_i}{2} + \frac{1}{4kT_i}\Big) 
g_{\mu\nu} (\tilde\nabla_\rho \tilde\nabla_\sigma h^{\rho\sigma} 
- \tilde\nabla^2 \tilde h) \Big]_{y=y_i}\,.
\eea
Then, from (\ref{eqn:gravonlyeq1}) and (\ref{eqn:gravonlyeq2}) 
we get the equations 
which the graviton Green's function should satisfy, 
whereas (\ref{eqn:gravonlybbeq}) provides the boundary conditions 
on the branes:
\bea\label{eqn:greenftneq1}
&& \tilde\nabla_\mu \tilde\nabla_\rho G^\rho_{\nu;\mu'\nu'} 
+ \tilde\nabla_\nu \tilde\nabla_\rho G^\rho_{\mu;\mu'\nu'} 
- \tilde\nabla^2 G_{\mu\nu;\mu'\nu'} - \tilde\nabla_\mu 
\tilde\nabla_\nu G^\rho_{\rho;\mu'\nu'} \nn\\
&&+ g_{\mu\nu} \Big(-\frac{2}{3}\tilde\nabla_\rho 
\tilde\nabla_\sigma G^{\rho\sigma}_{;\mu'\nu'} 
+ \frac{1}{2} \tilde\nabla^2 G^\rho_{\rho;\mu'\nu'} \Big) \nn\\
&& - G_{\mu\nu;\mu'\nu'}{}'' + g_{\mu\nu} \frac{1}{4} 
G^\rho_{\rho;\mu'\nu'}{}'' 
+ \frac{a'}{a} g_{\mu\nu} G^\rho_{\rho;\mu'\nu'}{}'
+ \frac{4a'{}^2}{a^2} G_{\mu\nu;\mu'\nu'} + \frac{H^2}{2a^2} 
g_{\mu\nu} G^\rho_{\rho;\mu'\nu'} \nn\\
&& + \frac{a}{3a'}\Big(\frac{1}{2}(\tn_\mu\tn_\rho 
G^\rho_{\nu;\mu'\nu'}{}' + \tn_\nu\tn_\rho G^\rho_{\mu;\mu'\nu'}{}') 
- \tn_\mu\tn_\nu G^\rho_{\rho;\mu'\nu'}{}' \Big) \\
&&+ g_{\mu\nu} \frac{a}{12a'}(\tn^2 G^\rho_{\rho;\mu'\nu'}{}' 
- \tn_\rho \tn_\sigma G^{\rho\sigma}_{;\mu'\nu'}{}') \nn\\
&&= \frac{1}{2M^3} (g_{\mu\mu'} g_{\nu\nu'} + g_{\mu\nu'} 
g_{\nu\mu'}) \frac{\delta^{(4)}(x-x')}{\sqrt{g}} \delta(y-y') \,,\nn\\
\nn\\
\label{eqn:greenftneq2}
&& -\tn_\mu\tn_\rho\tn_\sigma G^{\rho\sigma}_{;\mu'\nu'} 
+ \tn_\mu\tn^2 G^\rho_{\rho;\mu'\nu'} 
- \frac{a'}{a}\tn_\mu G^\rho_{\rho;\mu'\nu'}{}' 
- \frac{3H^2}{a^2}\tn_\mu G^\rho_{\rho;\mu'\nu'} 
+ \frac{4a'}{a}\tn_\nu G^\nu_{\mu;\mu'\nu'}{}' \nn\\
&&= 0 \,,\\
\nn\\
\label{eqn:greenftnbbeq}
&& \Big[\,\theta_i (G_{\mu\nu;\mu'\nu'}{}' 
- g_{\mu\nu} \frac{1}{4}G^\rho_{\rho;\mu'\nu'}{}') 
- \Big(\frac{\lambda_i H^2}{a^2} + 2k T_i \Big) G_{\mu\nu;\mu'\nu'} 
- \frac{H^2}{2a^2} \Big(\lambda_i + \frac{3}{2kT_i} \Big) 
g_{\mu\nu}G^\rho_{\rho;\mu'\nu'} \nn\\
&&\quad + \frac{\lambda_i}{2}(\tilde\nabla_\mu \tilde\nabla_\rho 
G^\rho_{\nu;\mu'\nu'} 
+ \tilde\nabla_\nu \tilde\nabla_\rho G^\rho_{\mu;\mu'\nu'} 
- \tilde\nabla^2 G_{\mu\nu;\mu'\nu'} 
- \tilde\nabla_\mu \tilde\nabla_\nu G^\rho_{\rho;\mu'\nu'}) \nn\\
&&\quad - \Big(\frac{\lambda_i}{2} + \frac{1}{4kT_i}\Big) 
g_{\mu\nu} (\tilde\nabla_\rho \tilde\nabla_\sigma G^{\rho\sigma}_{;\mu'\nu'} 
- \tilde\nabla^2 G^\rho_{\rho;\mu'\nu'}) \Big]_{y=y_i} \nn\\
&&= 0\,.
\eea
Now we follow \cite{D'Hoker} and \cite{Naqvi} to 
solve (\ref{eqn:greenftneq1})-(\ref{eqn:greenftnbbeq}). 
First we decompose $G_{\mu\nu;\mu'\nu'}$ using 
the five independent bitensor bases:
\be\label{eqn:Gdecomp1}
G_{\mu\nu;\mu'\nu'} = \sum_{i=1}^{5} T_{\mu\nu;\mu'\nu'}^{(i)} 
A^{(i)}(u,y,y')\,,
\ee
where $u = \cosh H\mu -1$ with $\mu$ the geodesic distance 
between $x$ and $x'$, and
\bea\label{eqn:tensorbases}
T^{(1)}_{\mu \nu;\mu' \nu'} &=& g_{\mu \nu} \ g_{\mu ' \nu'} \,, \nn\\
T^{(2)}_{\mu \nu;\mu' \nu'} &=& \pa_\mu u \pa_\nu u \pa_{\mu'} 
u \pa_{\nu'} u\,, \nn\\
T^{(3)}_{\mu \nu;\mu' \nu'} &=& \pa_\mu \pa_{\mu'} u \pa_\nu \pa_{\nu'} 
u + \pa_\mu \pa_{\nu'} u \pa_\nu \pa_{\mu'} u\,, \\ 
T^{(4)}_{\mu \nu;\mu' \nu'} &=& g_{\mu \nu} \pa_{\mu'} u \pa_{\nu'} 
u + g_{\mu ' \nu'} \pa_\mu u \pa_\nu u \,, \nn\\
T^{(5)}_{\mu \nu;\mu' \nu'} &=& \pa_\mu \pa_{\mu'} u \pa_\nu u \pa_{\nu'} 
u + \pa_\nu \pa_{\mu'} u \pa_\mu u \pa_{\nu'} u
 + (\mu' \leftrightarrow \nu') \,. \nn
\eea 
Then we reorganize (\ref{eqn:Gdecomp1}) into
\bea\label{eqn:greenftnansatz}
G_{\mu \nu ;\mu' \nu'}& =& T^{(3)} a^4 G_1(u,y,y') + T^{(1)} G_2(u,y,y') \nn\\
& & +\tn_\mu L_{\nu;\mu' \nu'} + \tn_\nu L_{\mu; \mu' \nu'} +
\tn_{\mu'} \Lambda_{\mu \nu;\nu'} + \tn_{\nu'} \Lambda_{\mu \nu;\mu'},
\eea
where
\bea
\Lambda_{\nu;\mu' \nu'} &=& g_{\mu' \nu'} \pa_\nu u \,a^2 A(u,y,y') 
+ \pa_{\mu'} u \pa_{\nu'} u \pa_\nu u \,a^4 C(u,y,y') \nn\\
&&+\pa_\nu(\pa_{\mu'} u \pa_{\nu'} u) \,a^4 B(u,y,y') \,,\\
\Lambda_{\mu \nu;\nu'} &=& g_{\mu \nu} \pa_{\nu'} u \,a^2 A(u,y,y') 
+ \pa_\mu u \pa_\nu u \pa_{\nu'} u \,a^4 C(u,y,y') \nn\\
&&+ \pa_{\nu'} (\pa_\mu u \pa_\nu u) \,a^4B(u,y,y') \,.
\eea
With this reorganization, we can single out the physically meaningful 
part, $G_1$ and $G_2$. 
The $A^{(i)}$'s and $G_1, G_2, A, B, C$ are related by
\bea
&&A^{(1)} = G_2 + 4H^2(1+u)A\,,\quad
A^{(2)} = 4a^4 \dot C\,,\quad
A^{(3)} = a^4(G_1 + 4B)\,, \nn\\
&&A^{(4)} = a^2 \bar f \equiv a^2 \{2\dot A + 2H^2 (2B + (1+u) C)  \} \,,\quad
A^{(5)} = 2a^4 (\dot B + C)\,,
\eea
with an overdot implying $\partial/\partial u$.
Upon plugging (\ref{eqn:greenftnansatz}) into (\ref{eqn:greenftneq1}), we get
\bea
&&T^{(1)} H^2 E_1 + T^{(2)} a^2 E_2 + T^{(3)} a^2 E_3 
+ g_{\mu \nu} \pa_{\mu'} u \pa_{\nu'} u \,H^2 E_4 
+ g_{\mu ' \nu'} \pa_\mu u \pa_\nu u \,\tilde E_4 + T^{(5)} a^2 E_5 \nn\\
&&= \frac{1}{2M^3} (g_{\mu\mu'} g_{\nu\nu'} + g_{\mu\nu'} 
g_{\nu\mu'}) \frac{\delta^{(4)}(x-x')}{\sqrt{g}} \delta(y-y') \nn\\
&&= \frac{1}{2M^3} \frac{a^4}{H^4} \Big( \frac{2T^{(2)}}{(2+u)^2} 
+ T^{(3)} - \frac{T^{(5)}}{2+u} \Big) 
\Big(\frac{H^4}{a^4} \frac{\delta(u)}{4\pi^2 u} \Big) \delta(y-y') \,, 
\eea
where we have converted $g_{\mu\mu'} g_{\nu\nu'} + g_{\mu\nu'} 
g_{\nu\mu'}$ into $T^{(i)}$'s 
applying the relations given in \cite{D'Hoker}. To convert 
$\delta^{(4)}(x-x')$ into $\delta(u)$, 
we start from a Euclideanized global coordinate system 
$(\tau, \rho, \theta, \phi)$
\be
\frac{H^2}{a^2}  ds^2 = \cosh^2 \rho \;d\tau^2 + d\rho^2 
+ \sinh^2 \rho (d\theta^2 + \sin^2\theta \;d\phi^2)\,, 
\ee
where $u$ measured from $(0,0,0,0)$ to $(\tau, \rho, 0, 0)$ 
can be explicitly written as
\be
u = \cosh \rho \cosh \tau -1\,.
\ee
Note that at a given $u$, $\tau$ is restricted by 
$-\cosh^{-1}(1+u)<\tau<\cosh^{-1}(1+u)$.
Transforming to $(\tau, u,\theta,\phi)$, 
\be
\frac{H^2}{a^2} ds^2 
&=& \frac{u(1+u)^2(2+u){\rm sech}^2\tau}{(1+u)^2-\cosh^2\tau}d\tau^2 
+ \frac{1}{(1+u)^2-\cosh^2\tau}du^2 \nn\\
&&- \frac{2(1+u)\tanh\tau}{(1+u)^2-\cosh^2\tau}d\tau du 
+ \frac{(1+u)^2-\cosh^2\tau}{\cosh^2\tau} (d\theta^2 + \sin^2\theta d\phi^2)\,.
\ee
Then, when integrated with a function depending on $u$ only,
\bea
\frac{\delta^{(4)}(x-x')}{\sqrt{g}} &=& 
\frac{H^4}{a^4} \frac{\delta(\tau)\delta(u)\delta(\theta)\delta(\phi)}
{(1+u){\rm sech}^2 \tau \sqrt{(1+u)^2 {\rm sech}^2\tau-1} \sin\theta} \nn\\
&=& \frac{H^4}{a^4} 
\frac{\delta(u)\delta(\tau)}{4\pi (1+u){\rm sech}^2 \tau \sqrt{(1+u)^2 
{\rm sech}^2\tau-1}} \nn\\
&=& \frac{H^4}{a^4} \frac{\delta(u)}{4\pi^2 u}  \,.
\eea
In the last step we use, for arbitrary function $f$ of $u$,
\be
\int du d\tau \sqrt{g(\tau,u)}
\frac{\delta(\tau)\delta(u)}{\sqrt{g(\tau,u)}}f(u) 
= f(0) = \int du d\tau \sqrt{g} \frac{\delta(u)}{\int d\tau \sqrt{g}} f(u)\,,
\ee
and then replace $\delta(\tau)/\sqrt{g}$ by $1/\int d\tau \sqrt{g}$, where
\bea
\int d\tau \sqrt{g} &=& \int_{-\cosh^{-1}(1+u)}^{\cosh^{-1}(1+u)}d\tau\; 
(1+u)^2 {\rm sech}^2\tau \sqrt{1-\frac{1}{(1+u)^2}-\tanh^2\tau} \nn\\
&=& \frac{\pi}{2}u(2+u)\,.
\eea
(\ref{eqn:greenftneq2}) gives
\be
g_{\mu' \nu'} \pa_\nu u \, E_5 + \pa_{\mu'} u \pa_{\nu'} 
u \pa_\nu u \, E_6 + \pa_\nu(\pa_{\mu'} u \pa_{\nu'} u) \,E_7 = 0\,.
\ee
Using several identities presented in \cite{D'Hoker},the
$E_i$'s can be worked out:
\bea
E_1 &=& \frac{-2{H^4} + 6{H^2}{{a'}^2}}{{a^2}}{G_1} 
+ \frac{2{H^4}}{3{a^2}}(1+u)\dot{G_1} + \Big(\frac{4a'}{a}{H^2} 
- \frac{{H^4}}{2a a'}\Big){G_1}' 
+ \frac{4{H^4}}{3{a^2}}u(2+u)\ddot{G_1} \nn\\
&&+ \frac{1}{2}{H^2}{G_1}'' - \frac{a}{6a'}\frac{{H^4}}{{a^2}}(1+u)\dot{G_1}' 
+ \frac{a}{6a'}\frac{{H^4}}{{a^2}}u(2+u)\ddot{G_1}' \nn\\
&&+ \frac{1}{{a^2}}u(2+u)\ddot{G_2} + \frac{1}{4a a'}u(2+u)\ddot{G_2}' \nn\\
&&+ \frac{8{H^2}}{{a^2}}{{(1+u)}^2}f 
+ \frac{{H^2}}{{a^2}}u(1+u)(2+u)\dot{f} \nn\\
&&+ \Big(-\frac{3{H^2}}{{a^2}}u(2+u) 
- \frac{6{H^2}}{{a^2}}{{(1+u)}^2} + \frac{4{H^2}}{{a^2}} 
+ \frac{3{{a'}^2}}{{a^2}}u(2+u)\Big)\bar{f} 
- \frac{2{H^2}}{{a^2}}u(1+u)(2+u)\dot{\bar{f}} \nn\\
&&+ \Big(\frac{a}{2a'}\frac{{H^2}}{{a^2}} 
+ \frac{2a'}{a}u(2+u) - \frac{a}{4a'}\frac{{H^2}}{{a^2}}u(2+u)\Big)\bar{f}' 
+ \frac{1}{4}u(2+u)\bar{f}'' 
- \frac{a}{4a'}\frac{{H^2}}{{a^2}}u(1+u)(2+u)\dot{\bar{f}}' \nn\\
&&+ \Big(\frac{24{H^2}}{{a^2}}{{(1+u)}^2} 
+ \frac{16{H^2}}{{a^2}}u(2+u)\Big)\dot{A} 
+ \frac{2a}{a'}\frac{{H^2}}{{a^2}}u(2+u)\dot{A}' \nn\\
&&+ \frac{8{H^2}}{{a^2}}u(1+u)(2+u)\ddot{A} 
+ \frac{a}{a'}\frac{{H^2}}{{a^2}}u(1+u)(2+u)\ddot{A}' \nn\\
&&+ \frac{-24{H^4} + 24{H^2}{{a'}^2}}{{a^2}}B 
+ \Big(\frac{16a'}{a}{H^2} - \frac{2a}{a'}\frac{{H^4}}{{a^2}}\Big)B' 
+ 2{H^2}B'' \nn\\
&&- \frac{8{H^4}}{3{a^2}}(1+u)\dot{B} 
- \frac{4a}{3a'}\frac{{H^4}}{{a^2}}(1+u)\dot{B}' \nn\\
&&- \frac{40{H^4}}{3{a^2}}(1+u)C 
- \frac{2a}{3a'}\frac{{H^4}}{{a^2}}(1+u)C\,' \,,
\\
E_2 &=& \ddot{Z} - \frac{10}{3}{H^2}\ddot{G_1} 
- \frac{2a}{3a'}{H^2}\ddot{G_1}' \nn\\
&&- \frac{8}{3}{H^2}(1+u)\stackrel{...}{B} 
+ \frac{8}{3}{H^2}\ddot{B} - \frac{4a}{3a'}{H^2}(1+u)\stackrel{...}{B}' 
+ \frac{4a}{3a'}{H^2}\ddot{B}' \nn\\
&&+ \frac{32}{3}{H^2}(1+u)\ddot{C} 
+ \Big(\frac{40}{3}{H^2} - 48{{a'}^2}\Big) \dot{C} - 4{a^2}\dot{C}\,'' \nn\\
&&+ \frac{16a}{3a'}{H^2}(1+u)\ddot{C}\,' 
+ \Big(\frac{32a}{3a'}{H^2} - \frac{32a'}{a}{a^2}\Big)\dot{C}\,' \,,
\eea
\bea 
E_3 &=& Z + \big(4{H^2} - 12{{a'}^2}\big){G_1} 
- \frac{4}{3}{H^2}(1+u)\dot{G_1} + \Big(-8a a' + \frac{a}{a'}{H^2}\Big){G_1}' 
- {H^2}u(2+u)\ddot{G_1} \nn\\
&&- {a^2}{G_1}'' + \frac{a}{3a'}{H^2}(1+u)\dot{G_1}' \nn\\
&&- 48{{a'}^2}B+\frac{16}{3}{H^2}(1+u)\dot{B} 
+ \frac{4}{3}{H^2}u(2+u)\ddot{B} + \Big(\frac{a}{a'}4{H^2} 
- \frac{32a'}{a}{a^2}\Big) B' \nn\\
&&- 4{a^2}B'' + \frac{8a}{3a'}{H^2}(1+u)\dot{B}' 
+ \frac{2a}{3a'}{H^2}u(2+u)\ddot{B}' \nn\\
&&+ \frac{8}{3}{H^2}(1+u)C - \frac{4}{3}{H^2}u(2+u)\dot{C} 
+ \frac{4a}{3a'}{H^2}(1+u)C\,' - \frac{2a}{3a'}{H^2}u(2+u)\dot{C}\,'  \,,
\\
E_4 &=& \big(-2{H^2}+6{{a'}^2}\big){G_1} - 2{H^2}(1+u)\dot{G_1} 
+ \Big(\frac{4a'}{a}{a^2} - \frac{a}{2a'}{H^2}\Big){G_1}' 
- \frac{4}{3}{H^2}\ddot{G_1} + \frac{1}{2}{a^2}{G_1}'' \nn\\
&&- \frac{a}{2a'}{H^2}(1+u)\dot{G_1}' - \frac{a}{6a'}{H^2}\ddot{G_1}' \nn\\
&&+ 10(1+u)\dot{f} + u(2+u)\ddot{f} + 8f \nn\\
&&+ 10\bar{f} + \frac{a}{2a'}\bar{f}' + 14(1+u)\dot{\bar{f}} 
+ \frac{a}{a'}(1+u)\dot{\bar{f}}' + 2u(2+u)\ddot{\bar{f}} 
+ \frac{a}{4a'}u(2+u)\ddot{\bar{f}}' \nn\\
&&+ \big(24{{a'}^2} - 72{H^2}\big)B + \Big(\frac{16a'}{a}{a^2} 
- \frac{2a}{a'}{H^2}\Big)B' 
+ 2{a^2}B'' + \big(-104{H^2}+24{{a'}^2}\big)(1+u)\dot{B} \nn\\
&&- \Big(\frac{8}{3}{H^2}{{(1+u)}^2} + \frac{40}{3}{H^2}u(2+u)\Big)\ddot{B} 
+ \Big(-\frac{6a}{a'}{H^2} + \frac{16a'}{a}{a^2}\Big)(1+u)\dot{B}' 
+ 2{a^2}(1+u)\dot{B}'' \nn\\
&&- \Big(\frac{2a}{3a'}{H^2}u(2+u) 
+ \frac{4a}{3a'}{H^2}{{(1+u)}^2}\Big)\ddot{B}' \nn\\
&&+ \big(-88{H^2} + 24{{a'}^2}\big)(1+u)C + \Big(\frac{16a'}{a}{a^2} 
- \frac{4a}{a'}{H^2}\Big)(1+u)C\,' + 2{a^2}(1+u)C\,'' \nn\\
&&+ \Big(-\frac{44}{3}{H^2}u(2+u) 
+ 12{{a'}^2}u(2+u) - \frac{160}{3}{H^2}{{(1+u)}^2}\Big)\dot{C} \nn\\
&&+ \Big(-\frac{8a}{3a'}{H^2}{{(1+u)}^2} - \frac{7a}{3a'}{H^2}u(2+u) 
+ \frac{8a'}{a}{a^2}u(2+u)\Big)\dot{C}\,' + {a^2}u(2+u)\dot{C}\,'' \nn\\
&&- 8{H^2}u(1+u)(2+u)\ddot{C} - \frac{a}{a'}{H^2}u(1+u)(2+u)\ddot{C}\,'  \,,
\\
\tilde E_4 &=& -\frac{10{H^4}}{3}\ddot{G_1} - \frac{2a}{3a'} H^4 \ddot{G_1}' 
- 4 \ddot{G_2} - \frac{a}{a'} \ddot{G_2}' 
+ 4{H^2} f + 2{H^2} (1+u)\dot{f} \nn\\
&&- 12 {a'}^2 \bar{f} + 2{H^2} (1+u)\dot{\bar{f}} + \Big(-\frac{8a'}{a}a^2 
+ \frac{a}{a'} H^2 \Big)\bar{f}' 
+ \frac{a}{a'} H^2 (1+u)\dot{\bar{f}}' - a^2 \bar{f}'' \nn\\
&&- 8{H^2} ((1+u)\ddot{A} + 2\dot{A}) - \frac{4a}{a'} H^2 ((1+u)\ddot{A}' 
+ 2\dot{A}') \nn\\
&&- \frac{8{H^4}}{3}\ddot{B} - \frac{4a}{3a'} H^4 \ddot{B}' 
+ \frac{8{H^4}}{3}\dot{C} + \frac{4a}{3a'} H^4 \dot{C}\,' \,,
\eea
\bea
E_5 &=& \dot{Z} + \frac{8}{3}{H^2}\dot{G_1} 
+ \frac{4}{3}{H^2}(1+u)\ddot{G_1} + \frac{a}{3a'}{H^2}\dot{G_1}' 
+ \frac{a}{6a'}{H^2}(1+u)\ddot{G_1}' \nn\\
&&+ \Big(\frac{16}{3}{H^2} - 24{{a'}^2}\Big)\dot{B} 
+ \frac{8}{3}{H^2}(1+u)\ddot{B} + \frac{2}{3}{H^2}u(2+u)\stackrel{...}{B} 
+ \Big(\frac{14a}{3a'}{H^2}-\frac{16a'}{a}{a^2}\Big)\dot{B}' \nn\\
&&- 2{a^2}\dot{B}'' + \frac{4a}{3a'}{H^2}(1+u)\ddot{B}' 
+ \frac{a}{3a'}{H^2}u(2+u)\stackrel{...}{B}' \nn\\
&&+ \Big(\frac{8}{3}{H^2} - 24{{a'}^2}\Big)C 
+ \frac{16}{3}{H^2}(1+u)\dot{C} - \frac{2}{3}{H^2}u(2+u)\ddot{C} 
+ \Big(\frac{10a}{3a'}{H^2} - \frac{16a'}{a}{a^2}\Big)C\,' \nn\\
&&- 2{a^2}C\,'' + \frac{8a}{3a'}{H^2}(1+u)\dot{C}\,' 
- \frac{a}{3a'}{H^2}u(2+u)\ddot{C}\,' \,,
\\
E_6 &=& -\big(6{H^4} + 4{H^2}a'{}^2\big)\dot{G_1} 
+ 10{H^4} (1+u)\ddot{G_1} + 2 {H^4} u(2+u)\stackrel{...}{G_1} 
- 2 a a'{H^2} {{\dot{G_1}}'} \nn\\
&&+ 18(1+u) \ddot{G_2} + 3 u (2+u)\stackrel{...}{G_2} \nn\\
&&- 36 (1+u)\big({H^2} - a'{}^2\big)\bar{f} 
- 3\big({H^2} \big(4 + 18u + 9{u^2}\big) 
- 2u(2+u)a'{}^2\big)\dot{\bar{f}} \nn\\
&&- 3 {H^2} u (1+u)(2+u)\ddot{\bar{f}}  
+ 18 a a'(1+u) {{\bar{f}}'} 
+ 3 a a'u(2+u) {{\dot{\bar{f}}}'} \nn\\
&&+ 144{H^2} (1+u) \dot{A} 
+ 36 {H^2} \big(2+6u+3{u^2}\big)\ddot{A}  
+ 12 {H^2} u(1+u)(2+u)\stackrel{...}{A} \nn\\
&&+ 16 {H^2} \big(-3{H^2} + a'{}^2\big) \dot{B}  
+ 8a a'{H^2} {{\dot{B}}'} \nn\\
&&+ \big(-24{H^4} + 32{H^2}a'{}^2\big) C  
- 24 {H^4} (1+u)\dot{C}  
+ 16a a'{H^2} {C\,'}  \,,
\\
E_7 &=& \big(-42 {H^2}-4  a'{}^2\big) \dot{G_1}
- 6 {H^2} (1+u)\ddot{G_1}  
- 2 {H^2}\stackrel{...}{G_1}  
- 2 a {a'} {{\dot{G_1}}'} \nn \\
&&+ 42 \dot{\bar{f}} 
+ 30 (1+u)\ddot{\bar{f}}  
+ 3u(2+u)\stackrel{...}{\bar{f}} \nn\\
&& + \big(-336  {H^2}+160  a'{}^2\big) \dot{B}
- (240 {H^2} - 16a'{}^2) (1+u) \ddot{B} 
-  24 {H^2} u(2+u)\stackrel{...}{B} \nn\\
&& + 80  a  {a'}  {{\dot{B}}'} 
 + 8 a {a'} (1+u) {{\ddot{B}}'} \nn\\
&& + \big(-168 {H^2} + 176  a'{}^2\big) C
- 8 (1+u) \big(51{H^2} - 28 a'{}^2\big) \dot{C} \nn\\
&& - 12 \big({H^2}  \big(10+26  u+13  {u^2}\big) 
- 2 u(2+u)  a'{}^2\big) \ddot{C}
 - 12 {H^2}u(1+u)(2+u) \stackrel{...}{C} \nn\\
&& + 88  a  {a'}  {C\,'} 
+ 112aa'(1+u) {{\dot{C}}\,'} 
+ 12aa' u (2+u) {{\ddot{C}}\,'} \,,
\\
E_8 &=& 36 \big(-{H^2}+a'{}^2\big) G_1  
+ 2 (1+u)  \big(-3  {H^2}+4 a'{}^2\big) \dot{G_1} \nn\\
&&- 2 {H^2} \ddot{G_1}   
+ 18  a  {a'}  {{G_1}'} 
+ 4  a a' (1+u) {{\dot{G_1}}'} \nn\\
&&+ 18 \bar{f} + 24 (1+u) \dot{\bar{f}} + 3 u (2+u) \ddot{\bar{f}} \nn\\
&&- 144 \big({H^2}-{{({a'})}^2}\big) B
- 16 (1+u) \big(12  {H^2}-7  a'{}^2\big) \dot{B} 
- 8  u(2+u) \big(3 {H^2}-2  a'{}^2\big) \ddot{B} \nn\\
&&+ 72  a  {a'}  B' 
+ 56  a a' (1+u)  {\dot{B}}' 
+ 8  aa'  u  (2+u)   {{\ddot{B}}'} \nn\\
&&- 8 (1+u)  \big(21  {H^2}-10 a'{}^2\big) C
- 8 \big(3 {H^2} \big(4+10u+5{u^2}\big) - u  (2+u) a'{}^2\big) \dot{C} \nn\\
&&- 12 {H^2} u(1+u)(2+u) \ddot{C}
+ 40  a a' (1+u)  {C\,'} 
+ 4  a a' u  (2+u) {{\dot{C}}\,'} \,,
\eea
with
\bea
f &\equiv& -2\dot A + 2H^2 (2B + (1+u) C) \,,\\
\label{eqn:Zformula}
Z &\equiv& 2f - 2\bar f - \frac{a}{a'}\bar f' = -8 \dot{A} 
- \frac{2a}{a'}(\dot{A}'+H^2(2B'+(1+u)C\,')) \,.
\eea
The brane-boundary part can be worked out similarly:
\bea
&&[T^{(1)} H^2 E_1^{\rm (bb)} + T^{(2)} a^2 E_2^{\rm (bb)} 
+ T^{(3)} a^2 E_3^{\rm (bb)} \nn\\
&&\quad + g_{\mu \nu} \pa_{\mu'} u \pa_{\nu'} u \,H^2 E_4^{\rm (bb)} 
+ g_{\mu ' \nu'} \pa_\mu u \pa_\nu u \, \tilde E_4^{\rm (bb)} 
+ T^{(5)} a^2 E_5^{\rm (bb)}]_{y=y_i} = 0 \,, 
\eea
where
\bea
E_1^{\rm(bb)} &=& -\Big({H^2} kT_i+\frac{{H^4}}{{a^2}} 
\Big(\frac{3}{k T_i}+4\lambda_i \Big)\Big)G_1+\frac{{H^4}}{{a^2}}
\Big(\frac{3}{2kT_i}
+2\lambda_i \Big)(1+u)\dot{G_1} \nn\\
&&+\frac{{H^4}}{{a^2}}\Big(\frac{1}{2k T_i}+\lambda_i 
\Big)u (2+u)\ddot{G_1}-\frac{\theta_i }{2}
{H^2}{{G_1}'}  \nn\\
&&-\frac{3}{{a^2}}\Big(\frac{1}{k T_i}+\lambda_i \Big)G_2
+\frac{3}{{a^2}}\Big(\frac{1}{k T_i}+\lambda_i \Big)(1+u)\dot{G_2}
+\frac{1}{{a^2}}\Big(\frac{3}{4kT_i}+\lambda_i \Big)u (2+u)\ddot{G_2}  \nn\\
&&+\frac{4\lambda_i  {H^2}}{{a^2}} {{(1+u)}^2}f
+\frac{\lambda_i  {H^2}}{2{a^2}} u (1+u) (2+u)\dot{f} \nn\\
&&-\Big(\frac{1}{2}k T_i u (2+u)+\frac{3{H^2}}{{a^2}} 
\big(1+6 u+3 {u^2}\big) \Big(\frac{1}{2kT_i}+\lambda_i \Big)\Big)\bar{f} \\
&&-\frac{3{H^2}}{2{a^2}}\Big(\frac{1}{2k T_i}
+\lambda_i \Big)u (1+u) (2+u)\dot{\bar{f}}
-\frac{\theta_i}{4} u (2+u)\bar{f}'  \nn\\
&&+\frac{6{H^2}}{{a^2}}\Big(\frac{1}{k T_i}+2 
\lambda_i \Big)\big(2+6 u+3 {u^2}\big)\dot{A}
+\frac{3{H^2}}{{a^2}}\Big(\frac{1}{kT_i}
+2\lambda_i \Big)u (1+u) (2+u)\ddot{A}  \nn\\
&&-4{H^2} \Big(k T_i+\frac{{H^2}}{{a^2}} 
\Big(\frac{3}{k T_i}+6\lambda_i \Big)\Big)B-2\theta_i H^2 B'
-\frac{6{H^4}}{{a^2}}\Big(\frac{1}{k T_i}+2\lambda_i \Big)(1+u)C  \,, \nn
\\
E_2^{\rm (bb)} &=& - \lambda_i {H^2}\ddot{G_1} + \lambda_i \ddot{f} 
+ 8kT_i  {a^2}\dot{C} + 4\theta_i   {a^2}\dot{C}\,' \,,
\\
E_3^{\rm (bb)} &=& \frac{\lambda_i}{2}\big(-{H^2}u(2+u)\ddot{G_1} 
- 2{H^2}(1+u)\dot{G_1} + 4{H^2}{G_1}\big) 
+ 2kT_i {a^2}{G_1} + \theta_i {a^2}{G_1}' 
+ \lambda_i f \nn\\
&&+ 8kT_i {a^2}B + 4\theta_i {a^2}B'  \,,
\\
E_4^{\rm (bb)} &=& -\Big(10\lambda_i H^2 + \frac{9H^2}{kT_i} 
+ a^2kT_i\Big)G_1 
- \Big(2\lambda_i H^2 + \frac{3H^2}{2kT_i}\Big)(1+u)\dot{G_1} \nn\\
&&- {H^2}\Big(\frac{1}{2kT_i} + \lambda_i \Big)\ddot{G_1} 
- \frac{\theta_i }{2}a^2G_1' \nn\\
&&+ 4\lambda_i f + 5\lambda_i (1+u)\dot{f} 
+ \frac{\lambda_i }{2}u(2+u)\ddot{f} \nn\\
&&+ 9\Big(\frac{1}{2kT_i} + \lambda_i \Big)\bar{f} 
+ 12\Big(\frac{1}{2kT_i} + \lambda_i \Big)(1+u)\dot{\bar{f}} 
+ \frac{3}{2}\Big(\frac{1}{2kT_i} + \lambda_i \Big)u(2+u)\ddot{\bar{f}} \nn\\
&&- \Big(72H^2\Big(\frac{1}{2kT_i} + \lambda_i \Big) 
+ 4a^2kT_i\Big)B - 2\theta_i a^2B' 
- \Big(96H^2\Big(\frac{1}{2kT_i} + \lambda_i \Big) 
+ 4a^2kT_i\Big)(1+u)\dot{B} \nn\\
&&- 2\theta_i a^2(1+u)\dot{B}' - 12{H^2}\Big(\frac{1}{2kT_i} 
+ \lambda_i \Big)u(2+u)\ddot{B} \nn\\
&&- \Big(84{H^2}\Big(\frac{1}{2kT_i} + \lambda_i \Big) 
+ 4{a^2}kT_i\Big)(1+u)C - 2\theta_i {a^2}(1+u)C\,' \nn\\
&&- \Big(12{H^2}\Big(\frac{1}{2kT_i} + \lambda_i \Big)\big(u(2+u) 
+ 4{{(1+u)}^2}\big) + 2{a^2}kT_i u(2+u)\Big)\dot{C} \nn\\
&&- 6{H^2}\Big(\frac{1}{2kT_i} + \lambda_i \Big)u(1+u)(2+u)\ddot{C} 
- \theta_i {a^2}u(2+u)\dot{C}\,' \,,
\eea
\bea
&&\hspace{-20pt} \tilde E_4{}^{\rm (bb)} =-\lambda_i {H^4}\ddot{G_1} 
- \lambda_i \ddot{G_2} + 2kT_i {a^2}\bar{f} 
+ \theta_i {a^2}\bar{f}' + 2\lambda_i {H^2}f 
+ \lambda_i {H^2}(1+u)\dot{f}  \,,\\
&&\hspace{-20pt} E_5^{\rm (bb)} =4kT_i {a^2}(\dot{B} + C) 
+ 2\theta_i{a^2}(\dot{B}' + C\,') 
+ \frac{\lambda_i }{2}\big({H^2}(1+u)\ddot{G_1} 
+ 2{H^2}\dot{G_1}\big) 
+ \lambda_i \dot{f} \,.
\eea
$E_2$, $E_3$ and $E_5$ give the bulk equation for $G_1$: 
when $u \neq 0$ or $y \neq y'$,
\bea\label{eqn:bulkG1eq}
&&\hspace{-20pt} \int_\infty^u du' \int_\infty^{u'} d u''  E_2 
+ 2\int_\infty^u du'  E_5 - E_3  \nn\\
&&\hspace{-20pt} = H^2  (2  G_1 + 4 (1+u)\dot{G_1} 
+ u (2+u) \ddot{G_1} ) + 12 a'{}^2 G_1 + 8 a a' G_1\,' + a^2 G_1\,'' = 0 \,,
\eea
and for $u = 0$ and $y = y'$,
\bea\label{eqn:bulkG1eqat0}
&&\hspace{-20pt} E_2 + 2 \dot E_5 - \ddot E_3  
= \ddot{\rm (\ref{eqn:bulkG1eq})}\nn\\
&&\hspace{-20pt} 
= \frac{1}{8\pi^2M^3a^2}\Big\{ \Big( \frac{1}{2u}
+\frac{1}{u^2}-\frac{2}{u^3} \Big) \delta(u) 
+ \Big( -\frac{1}{u}+\frac{2}{u^2} \Big) \dot\delta(u) 
- \frac{1}{u}\ddot\delta(u)  \Big\} \delta(y-y') \,.
\eea
Similarly from $E_2^{\rm (bb)}$, $E_3^{\rm (bb)}$ and $E_5^{\rm (bb)}$, we get
\bea\label{eqn:bbG1eq}
&&\hspace{-20pt} \int_\infty^u du' \int_\infty^{u'} d u''  E_2^{\rm (bb)} 
+ 2\int_\infty^u du' E_5^{\rm (bb)} - E_3^{\rm (bb)} \nn\\
&&\hspace{-20pt} = \Big[-2kT_ia^2 G_1 
+ \frac{\lambda_iH^2}{2} (4 (1+u) \dot{G_1} 
+ u  (2+u)\ddot{G_1} ) - \theta_i a^2 G_1\,' \Big]_{y=y_i} = 0\,.
\eea
Note that when performing integrations over $u$, 
we set the integration constants to be zero, because 
we want our solutions to die off as $u$ gets large. 
Assuming 
\be\label{eqn:G1ansatz}
G_1(u,y,y') = \int dp \;\frac{X(x;p)}{\sqrt{x^2-1}} 
\cdot (1-z^2)^2 Y(z,z';p) \,,
\eea
with $x=1+u$ and $z=\tanh k(y-y_0)$, (\ref{eqn:bulkG1eq}) is written
\bea
&&- \frac{1}{\sqrt{x^2-1}}\Big\{ (1-x^2) 
\frac{d^2 X}{dx^2} -2x \frac{d X}{dx} 
- \frac{1}{1-x^2}X \Big\} \cdot (1-z^2)^2 Y \nn\\
&&+ \frac{X}{\sqrt{x^2-1}} \cdot (1-z^2)^2 \Big\{(1-z^2) 
\frac{d^2 Y}{dz^2}  - 2z \frac{d Y}{dz} -  \frac{4}{1-z^2}Y \Big\} = 0\,.
\eea
The complete orthonormal basis for $X$ is given \cite{Grosche:1987de} by 
\be
X(x;p) = \frac{\Gamma(ip-1/2)}{\Gamma(ip)} P_{ip-1/2}^1(1+u)\,,\qquad p > 0\,,
\ee
whose eigenvalue is $p^2+\frac{1}{4}$, \ie,
\be\label{eqn:G1Xeq}
(1-x^2) \frac{d^2}{dx^2}X(x;p) -2x \frac{d}{dx}X(x;p) 
- \frac{1}{1-x^2}X(x;p) = \Big( p^2+\frac{1}{4} \Big) X(x;p)\,.
\ee
Note that these bases satisfy the boundary condition at $u\to\infty$ 
because $P_{ip-1/2}^1(1+u) \sim u^{ip-1/2}$ for large $u$. 
Then for any given $p$, the solution for $Y$ is
\bea\label{eqn:bulkG1Ysol}
Y(y,y';p)=A(z';p) P_{ip-1/2}^{-2}(z) + B(z';p) Q_{ip-1/2}^2(z) \,.
\eea
If we put the reference point $y'$ between $y=0$ and $y=L$, 
we actually have three copies of (\ref{eqn:bulkG1Ysol}):
\be
Y_i = A_i  P_{ip-1/2}^{-2}(z) + B_i Q_{ip-1/2}^2(z) \,, 
\qquad y\in {\cal I}_i\,,
\ee
where ${\cal I}_1 = (0, y')$, ${\cal I}_2 = (y', L)$ and ${\cal I}_3 
= (-L, 0)$. 
Also the correct interpretation of (\ref{eqn:bbG1eq}) is
\bea\label{eqn:bbG1eq2}
\Big\{-\frac{kv_i}{2}\Big( p^2+\frac{9}{4} \Big) (1-T_i^2)+2kT_i \Big\} Y(y_i) 
- \frac{1}{2} \{ \theta_i^+ Y'(y_i^+) + \theta_i^- Y'(y_i^-) \} = 0\,,
\eea
where $\theta_i^+ = -1$, $\theta_i^- = 1$
\footnote{In \cite{clp} $\theta_i$ was introduced to represent 
the ``outward" direction on the branes. That is, 
when we consider $0<y<L$ interval, on the $0$-brane ``outward" 
is in the direction of decreasing $y$
($\theta_0^+=-1$, where $+$ implies we are on the right side of the brane), 
whereas on the $L$-brane we will leave the interval by moving 
in the direction of increasing $y$($\theta_L^-=+1$). Then, 
if we are in the $-L<y<0$ interval, by the same reason we 
define $\theta_0^-\equiv+1$ and $\theta_{-L}^+=\theta_L^+\equiv-1$.}. 
Continuity of $Y$ at $y=0$, $L$ and $y'$ gives three equations 
for $A_i$'s and $B_i$'s: 
\be\label{eqn:Ybc1}
Y_1(0^+) = Y_3(0^-)\,,\quad 
Y_2(L^-) = Y_3(-L^+)\,, \quad 
Y_1(y'{}^-) = Y_2(y'{}^+)\,.
\ee
(\ref{eqn:bbG1eq2}) gives two more equations:
\bea\label{eqn:Ybc2}
Y_1{}'(0^+) - Y_3{}'(0^-) 
+ \{ k v_0 q (1-T_0^2) + 4 k T_0 \} \,Y_1(0) &=& 0\,, \nn\\
Y_3{}'(-L^+) - Y_2{}'(L^-) 
+ \{ k v_L q (1-T_L^2) + 4 k T_L \} \,Y_2(L) &=& 0\,, 
\eea
with $q = -p^2 - 9/4$. 
The integration of (\ref{eqn:bulkG1eqat0}) over $({y'}^-, {y'}^+)$ 
provides the last equation:
\bea\label{eqn:Ybc3}
&&\int_0^\infty dp \; \partial^2_u \Big( \frac{X}{\sqrt{u(2+u)}} \Big) 
\cdot Y' \Big|_{y=y'{}^-}^{y=y'{}^+} 
= \int_0^\infty dp \, \frac{\Gamma(ip-1/2)}{\Gamma(ip)} 
\frac{P_{ip-1/2}^3(1+u)}{\{ u(2+u) \}^{3/2}} \cdot Y' 
\Big|_{y=y'{}^-}^{y=y'{}^+}\nn\\
&& = \frac{\cosh^4 ky_0}{8\pi^2M^3} 
\Big\{ \Big( \frac{1}{2u}+\frac{1}{u^2}-\frac{2}{u^3} \Big) \delta(u) 
+ \Big( -\frac{1}{u}+\frac{2}{u^2} \Big) \dot\delta(u) 
- \frac{1}{u}\ddot\delta(u)  \Big\} \,.
\eea
Using,
\be
\Big|\frac{\Gamma(ip+1/2-\mu)}{\Gamma(ip)}\Big|^2 \int_1^\infty dx \, 
P_{ip-1/2}^\mu(x) P_{-ip'-1/2}^\mu(x) = \delta(p-p')\,,
\ee
and
\be
P_\nu^m(1+\epsilon) \approx \frac{\Gamma(\nu+m+1)}{m! \Gamma(\nu-m+1)} 
\Big( \frac{\epsilon}{2} \Big)^{m/2}\,,\quad m=0,1,2,\cdots\,,
\ee
(\ref{eqn:Ybc3}) becomes
\bea
&& \int_0^\infty dp \int_0^\infty du \frac{\Gamma(ip-1/2)}{\Gamma(ip)} 
P_{-ip'-1/2}^3(1+u) P_{ip-1/2}^3(1+u) 
\cdot Y' \Big|_{y=y'{}^-}^{y=y'{}^+} \nn\\
&&= \frac{\Gamma(ip'-1/2)}{\Gamma(ip')} \Big| 
\frac{\Gamma(ip')}{\Gamma(ip'-5/2)} \Big|^2 \{ Y_2'(y'{}^+;p') 
- Y_1'(y'{}^-;p') \} \nn\\
&& = \frac{\cosh^4 ky_0}{8\pi^2M^3} \Big[ \Big( \frac{1}{2u}
+\frac{1}{u^2}-\frac{2}{u^3} \Big) \{u(2+u)\}^{3/2} P_{-ip'-1/2}^3(1+u)  \nn\\
&&\qquad - \partial_u \Big\{ \Big( -\frac{1}{u}+\frac{2}{u^2} \Big) 
\{u(2+u)\}^{3/2} P_{-ip'-1/2}^3(1+u)  \Big\} \nn\\
&&\qquad + \partial_u^2 \Big( -\frac{1}{u} \{u(2+u)\}^{3/2} 
P_{-ip'-1/2}^3(1+u)  \Big) \Big]_{u\to0}  \nn\\
&& = \frac{\cosh^4 ky_0}{8\pi^2M^3} 
\Big(-\frac{\Gamma(-ip'+7/2)}{\Gamma(-ip'-5/2)}  \Big)\,,
\eea
or
\bea\label{eqn:G1norm}
Y_2'(y'{}^+;p') - Y_1'(y'{}^-;p') = \frac{\cosh^4 ky_0}{8\pi^2M^3} 
\Big(-\frac{\Gamma(-ip'+3/2)}{\Gamma(-ip')}  \Big)\,.
\eea
The general solutions for $A$'s and $B$'s with arbitrary $y$ and 
$y'$ are too lengthy to be written down. 
But since we are mainly interested in the gravity on the branes, 
we set $y=y'=0$, to get
\bea\label{eqn:completeG1sol}
G_1(u, y=y'=0) = \frac{1}{8\pi^2kM^3} \int_0^\infty dp\, {\cal N}(p) 
\frac{P_{ip-1/2}^1(1+u)}{\sqrt{u(2+u)}} \,,
\eea
where  
\bea\label{eqn:getN}
\hspace{-10pt} {\mathcal N}(p) &\equiv& 
- \frac{\Gamma(-ip+3/2)}{\Gamma(-ip)} \frac{\Gamma(ip-1/2)}{\Gamma(ip)} \cdot
\frac{b_L P_{ip-1/2}^{-2}(-T_0) 
- a_L Q_{ip-1/2}^2(-T_0)}{a_0 b_L - b_0 a_L} \nn\\
&=& p \tanh \pi p \;\frac{b_L P_{ip-1/2}^{-2}(-T_0) 
- a_L Q_{ip-1/2}^2(-T_0)}{a_0 b_L - b_0 a_L} \,.
\eea
$a_i$ and $b_i$ are given in (\ref{eqn:ourasandbs}) 
with $q$ and $\sqrt{9+4q}$ replaced by $-p^2-9/4$ and $2ip$ respectively.

We can decompose (\ref{eqn:completeG1sol}) into modes 
by finding the poles of (\ref{eqn:getN}), \ie,
\be\label{eqn:poles}
a_0 b_L - b_0 a_L =0\,. 
\ee
Once we choose our parameters from the allowed region, 
all the zeroes of (\ref{eqn:poles}) occur at non-negative $q$.

Let's perform contour integration to evaluate the integral 
in (\ref{eqn:completeG1sol}). 
Using \cite{Bateman} 3.3.1(8), 
\be
P_{ip-1/2}^1(1+u) = \frac{i}{\pi}\coth\pi p \;\{ Q_{ip-1/2}^1(1+u) 
- Q_{-ip-1/2}^1(1+u) \}\,,
\ee
(\ref{eqn:completeG1sol}) becomes
\bea\label{eqn:closetofinalG1sol}
G_1(u, y=y'=0) &=& \frac{1}{8\pi^2kM^3} \frac{i}{\pi} \int_0^\infty dp\, 
p \;\frac{b_L P_{ip-1/2}^{-2}(-T_0) - a_L Q_{ip-1/2}^2(-T_0)}{a_0 b_L 
- b_0 a_L} \nn\\ 
&&\hspace{80pt} \times \frac{Q_{ip-1/2}^1(1+u) 
- Q_{-ip-1/2}^1(1+u)}{\sqrt{u(2+u)}} \,.
\eea
With \cite{gradrhyz} 8.737.4,
\be
Q_{-\nu-1}^\mu (z) = \frac{\sin \pi(\nu+\mu)}{\sin \pi(\nu-\mu)} Q_\nu^\mu (z) 
- \frac{\pi\cos \pi\nu \cos \pi\mu }{\sin \pi(\nu-\mu)} P_\nu^\mu (z) \,,
\ee
we can show $\frac{b_L P_{ip-1/2}^{-2}(-T_0) 
- a_L Q_{ip-1/2}^2(-T_0)}{a_0 b_L - b_0 a_L}$ is even in $p$. 
Then
\bea\label{eqn:closertofinalG1sol}
\hspace{-20pt} &&G_1(u, y=y'=0) \nn\\
\hspace{-20pt} &&= \frac{1}{8\pi^2kM^3} 
\frac{i}{\pi} \Big\{ \int_0^\infty dp\, 
p \;\frac{b_L P_{ip-1/2}^{-2}(-T_0) - a_L 
Q_{ip-1/2}^2(-T_0)}{a_0 b_L - b_0 a_L} 
\frac{Q_{ip-1/2}^1(1+u)}{\sqrt{u(2+u)}} \nn\\
\hspace{-20pt} && \qquad - \int_0^{-\infty} d(-p)\, 
(-p) \;\frac{b_L P_{ip-1/2}^{-2}(-T_0) 
- a_L Q_{ip-1/2}^2(-T_0)}{a_0 b_L - b_0 a_L} 
\frac{Q_{ip-1/2}^1(1+u)}{\sqrt{u(2+u)}} \Big\}\nn\\
\hspace{-20pt} &&=\frac{1}{8\pi^2kM^3} \frac{i}{\pi} 
\int_{-\infty}^\infty dp\, 
p \;\frac{b_L P_{ip-1/2}^{-2}(-T_0) - a_L 
Q_{ip-1/2}^2(-T_0)}{a_0 b_L - b_0 a_L} 
\frac{Q_{ip-1/2}^1(1+u)}{\sqrt{u(2+u)}} \,.
\eea
Since $Q_{i p -1/2}^1(1+u) \sim e^{-ip H\mu}$ for large $p$, 
we close the contour below. Noting that poles occur at 
positive $q$'s, \ie, at pure imaginary $p$'s, we finally get
\bea\label{eqn:finalG1sol}
&&\hspace{-20pt} G_1(u, y=y'=0) = \sum_{j} G_1(u, y=y'=0;p_j) \nn\\
&&\hspace{-20pt} = \frac{1}{4\pi^2kM^3} \sum_{p_j>0} \Big[ p \frac{dp}{dq}\;
\frac{b_L P_{ip-1/2}^{-2}(-T_0) - a_L Q_{ip-1/2}^2(-T_0)}
{\partial_q(a_0 b_L - b_0 a_L)} \cdot 
\frac{Q_{ip-1/2}^1(1+u)}{\sqrt{u(2+u)}} \Big]_{p=-ip_j} \nn\\
&&\hspace{-20pt} = - \frac{1}{8\pi^2kM^3} \sum_{p_j>0} 
\frac{b_L P_{p_j-1/2}^{-2}(-T_0) - a_L Q_{p_j-1/2}^2(-T_0)}
{[\,\partial_q(a_0 b_L - b_0 a_L)\,]_{q=p_j^2-\frac{9}{4}}}  
\frac{Q_{p_j-1/2}^1(1+u)}{\sqrt{u(2+u)}}  \,.
\eea
Next, $E_2$ and $E_5$ give
\bea
&&\int_\infty^u d u'  E_5 - \int_\infty^u du' \int_\infty^{u'} d u''  E_2 \nn\\
&&=  \frac{2}{3}{H^2}u(2+u)(\ddot{B} 
- \dot{C})+4{H^2}(1+u)(\dot{B} - C) 
- \big(4{H^2}+24{{a'}^2}\big)(B - D) - 2{a^2}(B'' - D'') \nn\\
&&\quad + \frac{a}{3a'}{H^2}u(2+u)(\ddot{B}' 
- \dot{C}\,') + \frac{2a}{a'}{H^2}(1+u)(\dot{B}' - C\,') 
- \frac{16a}{a'}{{a'}^2}(B' - D') \nn\\
&&\quad + \frac{2}{3}{H^2}(2(1+u)\dot{G_1} + 7{G_1}) 
+ \frac{a}{6a'}{H^2}((1+u)\dot{G_1}' + 5{G_1}') \,,
\eea
with $D = \int_\infty^u du' C$. Solving it for $B-D$, we get 
\bea\label{eqn:B-Dsol}
B = D - \sum_{p_j} \frac{(1+u)\dot{G_1}(p_j)+3G_1(p_j)}{2q_j} \,.
\eea
Among the nine bulk and six brane-boundary equations we started with, 
we have solved two bulk and one brane-boundary ones 
to determine $G_1$ and $B-D$. 
Then we eliminate $\ddot{G_1}$, $G_1''$ and $B$'s 
from $\int\int du E_2$, $\tilde E_4$ and $E_7$ by 
(\ref{eqn:G1Xeq}) and (\ref{eqn:B-Dsol}), and solve 
them for $A$, $C$ and $G_2$ to get
\bea
D &=& \sum_{p_j}\frac{2\ddot{G_1}(p_j) + 6(1+u)\dot{G_1}(p_j) 
+ (18+3q_j)G_1(p_j)}{6(2+q_j)q_j}  \,,\\
B &=& \sum_{p_j}\frac{2\ddot{G_1}(p_j) - 3q_j(1+u)\dot{G_1}(p_j) 
- 6q_j G_1(p_j)}{6(2+q_j)q_j} \,,\\
\dot{A} &=& \sum_{p_j} \Big\{ \frac{2 {H^2}  
(-6+2  (-1+q_j)  u+(11+q_j)  {u^2}+12  {u^3}+3  {u^4})}{6 
(2+q_j)  {u^2}  {{(2+u)}^2}} G_1(p_j) \nn\\
&&\qquad +\frac{{H^2}  (1+u)  (16+q_j  u  
(-4+2  u+4  {u^2}+{u^3}))}{2q_j (2+q_j)  {u^2}  
{{(2+u)}^2}} \dot{G_1}(p_j) \Big\} \,,\\
\label{eqn:completeG2sol}
G_2 &=& -\sum_{p_j} \frac{2H^4}{3(2+q_j)^2} 
\Big\{ (q_j(2+q_j)+3(4+q_j)(1+u)^2) G_1(p_j) \nn\\
&&\hspace{80pt}+ 3u(1+u)(2+u) \dot{G_1}(p_j) \Big\}\,.
\eea
The last job is to verify the redundancy of the 
remaining four bulk equations and check 
if our solution satisfies the remaining five 
brane-boundary equations. It is easy to check 
\bea
\int d u  E_7  -  E_8  =  0\,,
\eea
upon getting rid of $\ddot G_1$ and $\stackrel{...}{G_1}$. 
As for $E_4$, we can replace $A$ by $Z$ using (\ref{eqn:Zformula}). 
The resulting equation has $B$, $C$, $G_1$ and $Z$. 
Then use $E_2$, $E_3$ and $E_5$ to 
write $Z$ in terms of $B$, $C$ and $G_1$. The last 
step is to use (\ref{eqn:B-Dsol}) and (\ref{eqn:G1Xeq}), 
and we will see $E_4 = 0$. Similarly, we can show 
that $E_1$ and $E_6$ are redundant.
Showing that our solution satisfies the brane-boundary 
equations is straightforward, using
(\ref{eqn:bbG1eq}) to get the final answer.

\end{document}